\documentclass[%
reprint,
superscriptaddress,
amsmath,amssymb,
aps,
prx,
floatfix,
]{revtex4-2}

\usepackage{graphicx}
\graphicspath{ {graphics/} }
\usepackage{hyperref}

\usepackage{qcircuit}
\usepackage{algorithm}
\usepackage{algpseudocode}

\usepackage{color}

\newcommand{\expect}[1]{\langle #1 \rangle}

\newcommand{\ket}[1]{| #1 \rangle}
\newcommand{\Exp}[1]{\mathrm{e}^{#1}}

\makeatletter
\@addtoreset{section}{part}
\def\@part[#1]#2{%
  \ifnum \c@secnumdepth >\m@ne
  \refstepcounter{part}%
  \addcontentsline{toc}{part}{\thepart\hspace{1em}#1}%
  \else
  \addcontentsline{toc}{part}{#1}%
  \fi
  {\parindent \z@ \raggedright
    \interlinepenalty \@M
    \normalfont\centering
    \ifnum \c@secnumdepth >\m@ne
    \Large\bfseries \partname
    \fi
    \huge \bfseries #2%
    \markboth{}{}\par}%
  \nobreak
  \vskip 3ex
  \@afterheading}
\renewcommand\partname{Supplemental Material}
\makeatother

\begin{document}

\title{Time-adaptive phase estimation}

\author{Brennan \surname{de Neeve}}
\email{brennan.mn@proton.me}
\affiliation{Institute for Quantum Electronics, ETH Z\"{u}rich, Otto-Stern-Weg 1, 8093 Z\"{u}rich, Switzerland}
\affiliation{Quantum Center, ETH Z\"{u}rich, 8093 Z\"{u}rich, Switzerland}

\author{Andrey V. Lebedev}
\affiliation{Theoretische Physik, Wolfgang-Pauli-Strasse 27, ETH Zürich, CH-8093, Zürich, Switzerland}
\affiliation{Current affiliation: Dukhov Research Institute of Automatics (VNIIA), Moscow, 127030, Russia}
\affiliation{Current affiliation: Advanced Mesoscience and Nanotechnology Centre, Moscow Institute of Physics and Technology (MIPT), Dolgoprudny, 141700, Russia}

\author{Vlad Negnevitsky}
\affiliation{Institute for Quantum Electronics, ETH Z\"{u}rich, Otto-Stern-Weg 1, 8093 Z\"{u}rich, Switzerland}

\author{Jonathan P. Home}
\email{jhome@phys.ethz.ch}
\affiliation{Institute for Quantum Electronics, ETH Z\"{u}rich, Otto-Stern-Weg 1, 8093 Z\"{u}rich, Switzerland}
\affiliation{Quantum Center, ETH Z\"{u}rich, 8093 Z\"{u}rich, Switzerland}

\begin{abstract}
  Phase estimation is known to be a robust method for single-qubit gate calibration in
  quantum computers \cite{2021russokirby}, while Bayesian estimation is widely used in
  devising optimal methods for learning in quantum systems \cite{2023gebhart}. We present
  Bayesian phase estimation methods that adaptively choose a control phase and the time of
  coherent evolution based on prior phase knowledge. In the presence of noise, we find
  near-optimal performance with respect to known theoretical bounds, and demonstrate some
  robustness of the estimates to noise that is not accounted for in the model of the
  estimator, making the methods suitable for calibrating operations in quantum
  computers. We determine the utility of control parameter values using functions of the
  prior probability of the phase that quantify expected knowledge gain either in terms of
  expected narrowing of the posterior or expected information gain. In particular, we find
  that by maximising the \emph{rate} of expected gain we obtain phase estimates having
  standard deviation a factor of $1.43$ larger than the Heisenberg limit using a classical
  sequential strategy. The methods provide optimal solutions accounting for available
  prior knowledge and experimental imperfections with minimal effort from the user. The
  effect of many types of noise can be specified in the model of the measurement
  probabilities, and the rate of knowledge gain can easily be adjusted to account for
  times included in the measurement sequence other than the coherent evolution leading to
  the unknown phase, such as times required for state preparation or readout.
\end{abstract}


\maketitle

\section{Introduction}

Phase estimation has found an increasing number of applications in metrology and quantum
computing in recent years. Although resources are considered differently in these two
settings \cite{2007vandam, 2009wiseman, 2014kaftal}, the methods of each have led to new
applications in the other. In metrology, classical averaging limits to estimates with
phase uncertainty $\Delta\hat{\phi}$ scaling at best according to the standard quantum
limit (SQL), $\Delta\hat{\phi} > 1/\sqrt{N}$, with the number of resources $N$, while
general strategies are fundamentally limited only by Heisenberg's uncertainty principle
and can obtain a $\sqrt{N}$ improvement over the SQL reaching the so-called Heisenberg
limit (HL), $\Delta\hat{\phi} > \pi/N$ \cite{2020gorecki}. Entanglement was initially
thought to be a key ingredient in schemes to reach Heisenberg \emph{scaling},
$\Delta\hat{\phi} \propto 1/N$ \cite{2004giovannetti}, but when it was also found that the
same scaling could be reached using sequentially prepared unentangled systems
\cite{2002luis, 2003rudolph, 2005deburgh, 2006giovannetti, 2009oloan, 2012boixo,
  2013maccone}\footnote{This is not necessarily true when noise is considered
  \cite{2014demkowicz-dobrzanski}.}, ideas from quantum computing \cite{1995kitaev,
  1996griffiths} soon led to new experimentally accessible metrology procedures with this
scaling \cite{2007higgins}. On the other hand, there have been several more recent
proposals to use metrology methods to calibrate operations for quantum computing
\cite{2009teklu, 2010brivio, 2015kimmel, 2019martinezgarcia, 2021russokirby}.

Many metrology proposals make use of a Bayesian approach to estimation which provides a
natural framework to describe \emph{adaptive} procedures, where the settings for future
experiments are modified based on the results of previous measurements \cite{1997wiseman,
  1998wiseman, 2000berry, 2001berry, 2005mitchell, 2007higgins, 2008boixo, 2009olivares,
  2023valeri, 2023smith}. While adaptive methods can lead to better performance, they are
generally more complex than non-adaptive strategies and can also be more difficult to
implement in some experiments. Remarkably, Higgins et al. found that they could reach
near-optimal Heisenberg scaling using a non-adaptive procedure by optimising the number of
measurements performed with different coherent applications of the unknown phase
\cite{2009higgins, 2009berry}.

While initial proposals like that of Higgins et al. \cite{2009higgins} considered ideal
settings with pure states and unitary operations, the role of noise and experimental
imperfections has been increasingly studied over time \cite{2011giovannetti}. This
development has been both in the understanding of the fundamental limits to precision
\cite{2007shaji, 2011escher, 2011maccone, 2012escher, 2012demkowicz-dobrzanski,
  2013kołodyński, 2014alipour, 2014macieszczak, 2017demkowicz-dobrzanski}, and in devising
better strategies to cope with non-ideal conditions, where the Bayesian framework with
adaptive measurements has proved useful \cite{2006cole, 2009maccone, 2009dorner,
  2010kolodynski, 2010kacprowicz, 2014vidrighin, 2020gutierrez-rubio}. In some cases,
adaptive methods have been used to design strategies better suited to the physics of
particular experiments. In recent developments for the sensing of magnetic fields with NV
centres, adaptive procedures were used to account for reduced visibility measurements
\cite{2012cappellaro, 2014hayes, 2016bonato, 2017bonato, 2019santagati, 2021joas,
  2021mcmichael, 2022zohar}\footnote{Readout can also be improved by adaptive methods
  \cite{2008myerson, 2016danjou}}. In many proposals the adaptive control acts on a phase
that can be seen as an adjustment of the measurement basis, but the time of coherent
interaction is chosen non-adaptively, e.g. as detailed in \cite{2009higgins}. Since the
interaction times proposed in \cite{2009higgins} are optimised without noise, and thus may
not be suited to experimental conditions, some experimenters perform optimisations using
numerical simulations that include relevant noise in order to find interaction times
better suited to their experiments\cite{2011said, 2012waldherr, 2012nusran, 2016bonato,
  2018danilin}. Recently, Belliardo and Giovannetti \cite{2020belliardo} have analytically
shown how to modify coherent interaction times in the non-adaptive procedure of Higgins et
al. \cite{2009higgins} to account for certain types of noise. Others have investigated the
possibility of also choosing the interaction time, or equivalently, in the case of some
optical measurements, the size of a so-called N00N-state \cite{2009berry}\footnote{This is
  an example where entanglement can be used to convert temporal resources into spacial
  resources \cite{2006giovannetti}.}, adaptively: initially in proposals without noise
\cite{2005mitchell, 2009berry}, or with mixed-state quantum computation \cite{2008boixo},
and later in proposals using numerical algorithms that can also account for noise and
imperfections \cite{2012granade, 2016wiebe, 2017paesani, 2021mcmichael, 2021gebhart}.

Commonly the performance of a strategy is only considered in the asymptotic regime, when
the number of resources (i.e. number of physical systems and total estimation time)
approaches infinity. Many experiments, however, could benefit from strategies that are
optimised for finite resources. Phase estimation in metrology is typically studied in two
settings. In one setting, often referred to as \emph{local}, the goal is to achieve
optimal sensitivity to phase fluctuations from an \textit{a priori} known phase
$\phi = \phi_0$. In the other setting, referred to as \emph{global}, it is assumed that
there is initially \emph{no a priori knowledge} of the phase \cite{2010kolodynski}. While
the amount of \emph{a priori} knowledge of the phase is not relevant in the asymptotic
regime, it can significantly change the optimal estimation strategies for finite
estimation times \cite{2011demkowicz-dobrzanski}. Therefore, it is useful to have methods
that can account for \emph{arbitrary a priori} phase knowledge.

In addition, the coherent interaction time in many proposals is optimised under the
assumption that this time dominates the experiment time. While this is true in the
asymptotic regime, many experiments may not reach this regime for all or at least a
significant portion of the estimation time \cite{2017bonato}. An extreme example is found
in GaAs quantum dots where the measurement time is several orders of magnitude longer than
the coherent evolution \cite{2011sergeevich}; here the authors show an exponential
improvement in the mean-square error (MSE) of parameter estimation by adaptively choosing
the time of coherent evolution.

In this paper we present time-adaptive phase estimation (TAPE), a method that allows for
the adaptive optimisation of the coherent interaction time and a control phase to be
adjusted to the resources of the experiment. TAPE can provide strategies optimised both
when the experiment time is proportional to the time of coherent interaction, and when the
experiment time is proportional to the number of measurements; the method also allows for
any resource allocation in between these two extreme cases. In addition TAPE can provide
optimal strategies for arbitrary \textit{a priori} knowledge of the phase, because the
choice of measurement settings depends only on the prior knowledge of the phase after the
last measurement, and not on any record of previous measurements. In contrast to earlier
works investigating similar adaptive procedures \cite{2005mitchell, 2009berry}, we use a
more general form for the measurement probabilities so that we can account for many types
of noise or imperfections.

We propose and analyse several different objective functions for the adaptive parameter
selection that lead to near-optimal performance with respect to known theoretical bounds,
with and without noise. When the experiment time is proportional to the time of coherent
interaction we reach uncertainties in the phase estimates that are a factor of $1.43$
larger than the HL. In addition, we find that TAPE is quite robust to errors that are not
accounted for in the model of the estimator. This demonstrates that, similarly to
proposals like robust phase estimation \cite{2021russokirby}, our methods are well-suited
to calibration of single-qubit operations in the context of quantum computing.

TAPE uses a numerical representation of the phase knowledge as a Fourier series that can
easily describe arbitrary prior knowledge, at the possible expense of increased
computation. Since the complexity of the Fourier representation (i.e. number of
coefficients to track in memory) increases with phase knowledge, the computation time
required for adaptive control of experimental parameters may become too large for some
practical applications. To tackle this limitation we propose a method that reduces the
interval over which the Fourier representation of the phase is used. This can
significantly reduce the required memory and time of computation.

\section{\label{sec:bayes-est}Bayesian estimation}

\begin{figure}[h]
  \centering
  \scalebox{1.0}{
  \Qcircuit @C=1em @R=.7em {
  \lstick{\ket{0}}    &\gate{H} &\ctrl{1}   &\gate{R_z(-\alpha)} &\gate{H} &\meter &\cw \\
  \lstick{\ket{\phi}} &\qw      &\gate{U^{k}} &\qw               &\qw      &\qw    &\qw
                                                                                   }}
  \caption{Quantum circuit for sequential phase estimation.}
  \label{fig:seq-phase-est-circuit}
\end{figure}
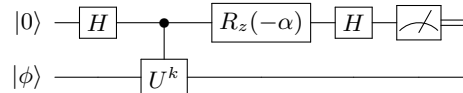

The evolution and measurement for a single step $s$ of sequential phase estimation can be
described by the quantum circuit in Fig.~\ref{fig:seq-phase-est-circuit}
\cite{2010nielsen}
where $H$ is the Hadamard gate, $R_z(-\alpha) = \Exp{i\alpha Z/2}$ with Pauli operator $Z$
is a z-rotation, and $\alpha \, \in [0, \pi]$ is a control phase to be
optimised. $U\ket{\phi} = \Exp{i\phi}\ket{\phi}$ with $\ket{\phi} \in \mathbb{C}^{2^n}$ an
eigenstate of the unitary $U \in \mathbb{C}^{2^n \times 2^n}$ with unknown phase
$\phi \, \in [0, 2\pi]$.  $k \in \mathbb{N}^+$ determines the number of applications of
the unitary evolution. The parameters $\alpha$ and $k$ will be chosen adaptively for each
measurement as detailed in section \ref{sec:adaptive-proc}. We denote the possible
measurement outcomes at step $s$ by $\xi \in \{\pm 1\}$, and in order to describe a range
of noise or measurement errors \cite{2017bonato}, we assume the outcome $\xi$ to occur
with probability
\begin{align}
  &P_{\xi}(\alpha, k\phi) = \nonumber \\
  &\frac{1}{2} \Big( 1 + \xi \big( (1-\lambda_k)
    + \lambda_k \zeta_k \cos\left(\alpha - k\phi\right) \big)
    \Big) \, , \label{eq:meas-prob}
\end{align}
where $\lambda_k, \zeta_k \in [0,1]$ allow respectively for the description of asymmetry
and reduced contrast in the measurement\footnote{While $\lambda_k < 1$ describes a reduced
  probability for the outcome $\xi = -1$, the reverse situation can always be described by
  relabelling the outcomes.}. The subscript $k$ of $\lambda_k, \zeta_k$ indicates that
these values can generally depend on $k$. Since the outcome $\xi$ and the control
parameters $\alpha$ and $k$ depend on $s$ we should really denote them
$\xi_s, \alpha_s, k_s$ but we omit the subscripts for simplicity of notation.

As in \cite{2005mitchell, 2009berry, 2012cappellaro} we use a Fourier series to write the
probability density describing our knowledge of the phase $\phi$ at step $s$
\begin{equation}
  p_s(\phi) = \sum_{n = -\infty}^\infty c_n^{(s)} \Exp{in\phi} \, . \label{eq:density}
\end{equation}
Given prior knowledge $p_{s-1}(\phi)$ and measurement outcome $\xi$, we update our
state of knowledge using Bayes' theorem
\begin{align}
  &p_s(\phi | \xi ; \alpha, k) \nonumber \\
  &= \frac{P_{\xi}(\alpha , k\phi) p_{s-1}(\phi)}{\Pi_{\xi}(\alpha,\, k)} \nonumber \\
  &\propto P_{\xi}(\alpha , k\phi) p_{s-1}(\phi) \nonumber \\
  &= \sum_{n = -\infty}^\infty \Bigg[
    \frac{1}{2}\big(1 + \xi(1 - \lambda_k)\big) c_n^{(s-1)} \nonumber \\
  &+ \xi\lambda_k\frac{\zeta_k}{4}
    \left( \Exp{i\alpha}c_{n+k}^{(s-1)} + \Exp{-i\alpha}c_{n-k}^{(s-1)} \right)
    \Bigg] \Exp{i n\phi} \, , \label{eq:update-rule}
\end{align}
where
\[
  \Pi_{\xi}(\alpha,k) = \int_0^{2\pi}\frac{d\phi}{2\pi} P_{\xi}(\alpha, k\phi) p_{s-1}(\phi)
\]
is the \emph{posterior} probability of outcome $\xi$. Equation \eqref{eq:update-rule}
specifies how to modify the coefficients $c_n^{(s-1)} \rightarrow c_n^{(s)}$ given the
measurement outcome.

We choose to use the estimator
\begin{equation}
  \hat{\phi} = \arg \int_0^{2\pi} \frac{d\phi}{2\pi} p_s(\phi) \Exp{i\phi}
                 = \arg\left(c_{-1}^{(s)}\right) \, .
               \label{eq:phase-est}
\end{equation}
A nice feature of the Fourier series representation for $p_s(\phi)$ is that the estimator
\eqref{eq:phase-est} depends only on one coefficient. Given an \emph{estimate}
$\hat{\phi}$ of $\phi$ (in general now, not necessarily given by \eqref{eq:phase-est}), we
quantify the uncertainty in $\hat{\phi}$ by the square root of the Holevo variance
\cite{2011holevo}:
\begin{align}
  \Delta\hat{\phi} = \sqrt{V(\hat{\phi})} = \sqrt{S(\hat{\phi})^{-2} - 1} \, ,
  \label{eq:Holevo-var}
\end{align}
where $S(\hat{\phi}) = |\expect{\Exp{i\hat{\phi}}} |$ is the \emph{sharpness} and the
angular brackets indicate an average over the estimates $\hat{\phi}$. If the estimate is
biased, one can use $S = \expect{\cos(\hat{\phi} - \phi)}$ instead, where $\phi$ is the
value of the correct system phase \cite{2009berry}. In \cite{2001berry} the authors show
that the estimator \eqref{eq:phase-est} minimises the Holevo variance
\eqref{eq:Holevo-var}.

\section{\label{sec:adaptive-proc}Adaptive procedures}

In the previous section we have described how in general to use Bayes' theorem to update
our knowledge of the phase $\phi$ given the measurement outcomes, and how to obtain an
estimate $\hat{\phi}$ from the density $p_s(\phi)$ at step $s$. The task of achieving
minimal uncertainty in the estimates obtained from \eqref{eq:phase-est} is then dependent
on the choices of the values of the control phase $\alpha$ and the number $k$ of
applications of $U$. In a sequential procedure, the state coherently evolves according to
$U^k$ for a time proportional to $k$, so that the choice of $k$ corresponds to choosing
the time of coherent evolution. In the following we will assume such a sequential
procedure, but we note that the results can also be applied in the case of some parallel
procedures by using entanglement.

Given prior knowledge $p_{s-1}(\phi)$ at step $s-1$ we look to choose the optimal control
phase for a given value of $k$ by maximising a function of $p_{s-1}(\phi)$, $k$, and
$\alpha$ that quantifies the \emph{expected knowledge gain} from the next measurement. One
might then expect that the optimal choice of $k$ can be determined by maximising the
expected knowledge gain over all possible values of $k$. This can be a good choice when
the evolution time of $U^k$ is negligible compared to other times in the experiment, as in
the example of GaAs quantum dots mentioned above \cite{2011sergeevich}. But in general
experiments with different values of $k$ require different resources in terms of execution
time, so that the expected knowledge gains for different values of $k$ are not directly
comparable\footnote{Even more generally, which resources are valuable depends on the
  setting and what the experimenter wants to optimise. e.g. they may have a restricted
  number of qubits to measure, so that number of measurements becomes the relevant
  resource. In this sense it is useful to have a method where the optimisation can be
  adjusted by the experimenter to describe best how they value their resources.}.

In \cite{2005mitchell, 2009berry}, the authors studied a method where the value of $k$ is
chosen adaptively. In \cite{2005mitchell}, $k$ corresponds to the number of photons in a
N00N state, however as shown for the noiseless case in \cite{2009berry} the sequential
procedure we discuss here is mathematically equivalent. This equivalence also holds under
certain types of noise \cite{2012boixo, 2013maccone}. Physically these two procedures are
different but they are connected by the use of entanglement to convert between temporal
and spatial resources \cite{2006giovannetti}.

In the case that the experiment time is proportional to $k$, the relevant resource is the
number of applications of $U$. This way of considering resources, which is typically
chosen in quantum metrology, is considered in \cite{2005mitchell, 2009berry}. In these
works the authors calculate the expected \emph{(differential) entropy} \footnote{The
  differential entropy is the entropy of a continuous random variable, but it lacks some
  important properties of the Shannon entropy for discrete random variables. See e.g.
  \cite{2005cover}, chapter 8. In this manuscript we will usually write simply ``entropy''
  when referring to the differential entropy.} of the posterior after the next measurement
using a Gaussian approximation. Motivated by the resource dependence on $k$ they choose
the value of $k$ that minimises the expected entropy \emph{divided by} $\ln(N)$. Here $N$
refers to the total number of resources used since the beginning of the estimation
sequence, where an initially uniform prior is assumed (no prior knowledge of the
phase). Since this method requires knowing $N$ assuming an initially uniform prior and
because the priors in the later stages of the estimation must be approximately Gaussian,
it is not well suited to incorporating arbitrary prior knowledge. Moreover, the approach
of dividing the entropy by $\ln(N)$ is based on scaling of the uncertainty in the phase
that is only valid without noise.

Here we study two different methods for adaptively choosing $k$ that have the following
features:
\begin{enumerate}
\item The resource dependence on $k$ is adjustable so that it can be chosen to match a
  given experiment. Since we consider sequential procedures, we take the resource
  requirement for performing an experiment with a particular value of $k$ to be the time
  required to perform that experiment.
\item The choice of $k$ for a given measurement is determined only by the prior
  probability density at that step of the estimation sequence.
\item The methods can be applied to different functions quantifying the \emph{expected
    knowledge gain} associated with a particular choice of $k$. We study two such possible
  functions possessing different benefits: the \emph{expected sharpness gain} and the
  \emph{expected (differential) entropy gain}.
\end{enumerate}
In the following we show how to calculate these two functions quantifying expected
knowledge gain exactly; with the exact expressions in hand these functions can be computed
for any prior, not only Gaussian. The expected knowledge gains are calculated from
expressions that can describe experiments with noise. This allows the methods to determine
good phase estimation procedures for noisy experiments, as well as in the noise-free case.

\subsection{\label{sec:know-gain} Expected knowledge gain}
Given prior knowledge $p_{s-1}(\phi)$ at step $s-1$, one might expect that in order to
minimise the Holevo variance of the \emph{estimates} $\hat{\phi}$ (equation
\eqref{eq:Holevo-var}) at step $s$, a good strategy can be to choose the control phase
$\alpha$ that minimises the \emph{expected} Holevo variance of the \emph{posterior
  Bayesian probability density} for the next measurement. However, it is shown in
\cite{2001berry, 2009berry} that in order to minimise the Holevo variance of the
\emph{estimates} $\hat{\phi}$ one should choose $\alpha$ to maximise the expected
\emph{sharpness} of the posterior Bayesian probability density for the next measurement
\begin{align*}
  \sum_{\xi} \Pi_{\xi}(\alpha, k\phi) S\left[p_s(\phi|\xi;\alpha, k)\right] \, ,
\end{align*}
where we denote the sharpness of $p_s(\phi)$ by
\begin{align}
  S\left[p_s(\phi)\right]
  &= \left| \int_0^{2\pi} \frac{d\phi}{2\pi} p_s(\phi) \Exp{i\phi} \right|
    = \left| c_{-1}^{(s)} \right| \, .
    \label{eq:sharpness}
\end{align}
Similarly to the estimate \eqref{eq:phase-est}, a nice feature of the Fourier series
representation for $p_s(\phi)$ is that the sharpness \eqref{eq:sharpness} depends only on
a single Fourier coefficient. Maximising the expected sharpness of $p_s(\phi)$ for the
next measurement is a good strategy since the sharpness of the estimates can be written as
the average of $S\left[p_s(\phi)\right]$ over possible measurement records; maximising the
sharpness of the estimates is equivalent to minimising the Holevo variance of the
estimates \cite{2009berry}. We note that the average of $S\left[p_s(\phi)\right]$ over
possible measurement records considered in \cite{2009berry} assumes a uniform prior at the
beginning of the estimation sequence, and this strategy may be less optimal for other
priors.


In order to evaluate the best strategy for the measurement at step $s$, we define the
\emph{expected sharpness gain} as
\begin{align}
  &\Delta_s S(\alpha,\, k) \equiv \nonumber \\
  &\sum_{\xi} \Pi_{\xi}(\alpha,\, k)
    \Big( S\left[ p_s(\phi|\xi ; \alpha, k) \right]
    - S\left[ p_{s-1}(\phi) \right] \Big) \nonumber \\
  &= - \left| c_{-1}^{(s-1)} \right| + \sum_{\xi} \bigg|
    \frac{1}{2}\big(1 + \xi(1 - \lambda_k)\big) c_{-1}^{(s-1)} \nonumber \\
  &\qquad\qquad\qquad + \xi\lambda_k\frac{\zeta_k}{4}
    \left( \Exp{i\alpha}c_{-1+k}^{(s-1)} + \Exp{-i\alpha}c_{-1-k}^{(s-1)} \right)
    \bigg| \, , \label{eq:sharpness-gain}
\end{align}
rather than working with the expected sharpness of the posterior directly. The reason for
working with the gain will become clear later on when we consider gain
\emph{rates}. Similarly, we define the \emph{expected entropy gain} for the probability
density $p_s(\phi)$ at step $s$,
\begin{align}
  &\Delta_s H(\alpha,\, k) \equiv \nonumber \\
  &\sum_{\xi} \Pi_{\xi}(\alpha,\, k)
    \Big( H[p_{s-1}(\phi)] - H[p_s(\phi | \xi ; \alpha, k)] \Big) \, ,
    \label{eq:entropy-gain-def}
\end{align}
where $H[p(\phi)]$ is the differential entropy of $p(\phi)$,
\begin{align*}
  H[p(\phi)] &= -\int_0^{2\pi} \frac{d\phi}{2\pi} p(\phi) \ln\left(\frac{p(\phi)}{2\pi}\right) \, .
\end{align*}

We note that $\Delta_s H(\alpha,\, k)$ can be seen as the \emph{expected information gain}
from the next measurement. It can be rewritten as (see Supplemental Material, section
\ref{sec:sup-ent-gain})
\begin{align*}
  \Delta_s H(\alpha,\, k)
  &= \sum_{\xi} \Pi_{\xi}(\alpha, k) D_{KL}[p_s\| p_{s-1}] \, ,
\end{align*}
where we denoted $p_{s-1} = p_{s-1}(\phi)$, $p_s = p_s(\phi | \xi ; \alpha, k)$ for short,
and $D_{KL}[p_s \| p_{s-1}]$ is the Kullback-Leibler divergence of the posterior from the
prior \cite{1951kullback, 1978kullback}. In ref. \cite{2020gutierrez-rubio} the authors
show that maximising expected entropy gain maximises the likelihood of estimating the true
phase. In terms of the coefficients of the prior, this can be computed by an expression of
the form
\begin{align}
  \Delta_s H(\alpha,\, k) &= f(\lambda_k, \zeta_k)
                            - \sum_{\xi} \Pi_{\xi}(\alpha,\, k)
                            \ln \left(\Pi_{\xi}(\alpha,\, k)\right) \nonumber \\
                          &+ \sum_{m=1}^\infty \Big( A_m \mathfrak{Re}
                            \left\{ \Exp{i 2m \alpha} c_{2mk}^{(s-1)} \right\} \nonumber \\
                          &+ B_m \mathfrak{Re}
                            \left\{ \Exp{i (2m-1) \alpha} c_{(2m-1)k}^{(s-1)} \right\}
                            \Big) \, . \label{eq:entropy-gain}
\end{align}
The $m$-dependent coefficients $A_m$ and $B_m$ are generally also functions of $\lambda_k$
and $\zeta_k$. Since our knowledge $p(\phi)$ after a finite number of measurements is
always described by a finite Fourier series, the sum over $m$ is always finite. The
derivation of this expression as well as the exact expressions for $f$, $A_m$, and $B_m$
are given in the Supplemental Material, section \ref{sec:sup-ent-gain}.

\begin{figure}[h]
  \centering
  \includegraphics[width=0.48\textwidth]{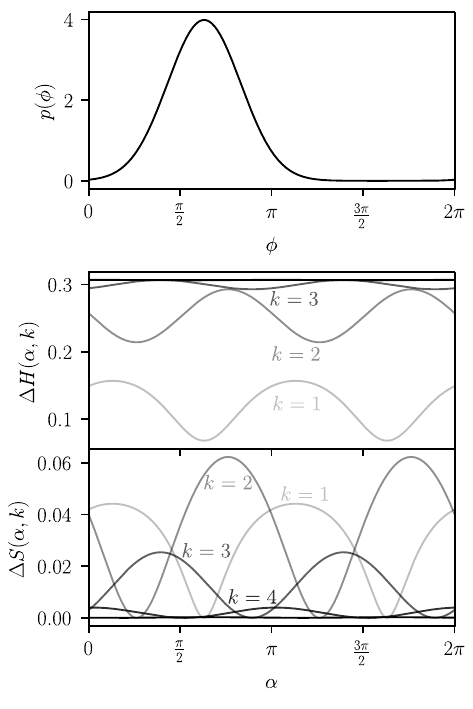}
  \caption{An example prior (top) is plotted along with the expected entropy gain,
    $\Delta H(\alpha, k)$ (middle) and expected sharpness gain $\Delta S(\alpha, k)$
    (bottom) as a function of $\alpha$ for $k \le 5$. $\Delta H(\alpha, k)$ and
    $\Delta S(\alpha, k)$ have a period of $\pi$ in $\alpha$ (if we see $\alpha$ as a
    change of the measurement basis, then $\alpha = \pi$ corresponds to the change
    $\ket{0} \rightarrow \ket{1}$, $\ket{1} \rightarrow \ket{0}$). The expected entropy
    gain increases to just over $0.3$ for large $k$, while the expected sharpness gain
    increases from $k = 1$ to $2$, and then decreases, reaching zero for $k > 4$. The
    gains for $k > 1$ do not align with $p(\phi)$ since the control phase $\alpha$ is
    applied once per measurement, whereas the unknown phase $\phi$ is applied $k$ times.}
  \label{fig:pior-and-gains-1-peak}
\end{figure}

Before discussing in detail how to use the expected gains to choose $k$ we discuss the
choice of the control phase $\alpha$ and compare the benefits of using sharpness to
quantify expected knowledge gain versus entropy. For both methods described in section
\ref{sec:choose-k} below, computing the expected gain for a given value of $k$ always
involves an optimisation of the control phase $\alpha$ for that particular $k$-value. An
example prior is shown in figure \ref{fig:pior-and-gains-1-peak} and the values of the
expected entropy and sharpness gains for the next measurement are plotted for $k \le
5$. The expected entropy gain increases significantly from $k=1$ to $k=2$ while the
increase for larger $k$-values is smaller. The expected sharpness gain also increases
significantly from $k=1$ to $k=2$, but contrary to the expected entropy gain, decreases
for $k>2$, and becomes zero for $k > 4$. This is because as $k$ increases, it is
increasingly likely that the posterior will have multiple peaks. When $k > 4$ the effect
of the measurement leads to a posterior density that is split into multiple peaks with an
envelope similar to the prior density, so that the sharpness is unchanged. Conversely, the
entropy still increases when the probability density is split into multiple peaks, so that
the expected entropy gain is large for all values of $k$.

\begin{figure}[h]
  \centering
  \includegraphics[width=0.48\textwidth]{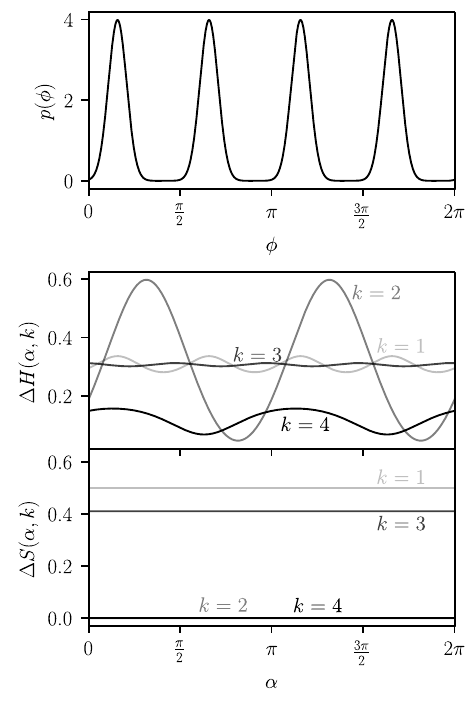}
  \caption{An example prior with four equally-spaced peaks (top) is plotted along with the
    expected entropy gain, $\Delta H(\alpha, k)$ (middle) and expected sharpness gain
    $\Delta S(\alpha, k)$ (bottom) as a function of $\alpha$ for $k \le 4$.
    $\Delta H(\alpha, k)$ quantifies three different strategies: sharpening the envelope
    of $p(\phi)$ ($k=1,3$), eliminating two peaks ($k=2$), and sharpening individual peaks
    ($k=4$). $\Delta S(\alpha, k)$ quantifies only strategies that sharpen the envelope
    ($k=1,3$) and $\Delta S(\alpha, k) = 0$ for $k = 2, 4$. $\Delta S(\alpha, k)$ is
    independent of $\alpha$ because the envelope of the prior is flat.}
  \label{fig:pior-and-gains-4-peaks}
\end{figure}

These differences between the entropy and the sharpness make them useful for different
purposes. Maximising the expected sharpness gain is useful to ensure that the probability
density $p(\phi)$ has a single peak -- this is important for obtaining an accurate
estimate of the phase. On the other hand, the entropy allows to quantify good strategies
when there are multiple peaks in $p(\phi)$. In figure \ref{fig:pior-and-gains-4-peaks} we
plot a prior with four peaks along with the expected entropy and sharpness gains to
illustrate these differences. We see that the maximum entropy gain corresponds to a
measurement with $k = 2$ that will eliminate two of the peaks in $p(\phi)$ with high
probability. In contrast, the expected sharpness gain for $k = 2$ is zero because the
density with either four or two equally spaced peaks has zero sharpness, meaning that the
strategy of eliminating two peaks is not quantified by the sharpness. The expected
sharpness gain in this case is non-zero only for odd values of $k$ since they will lead to
a narrowing of the envelope of $p(\phi)$. The entropy quantifies strategies that sharpen
the overall envelope ($k = 1, 3$), the strategy of eliminating two peaks ($k = 2$), and
the strategy of sharpening the individual peaks ($k = 4$).

\subsection{\label{sec:choose-k}Choosing $k$}
We now describe the two methods for choosing $k$. As noted above, if experiments with
different values of $k$ require different amounts of resources, comparing the expected
knowledge gain for different $k$ does not directly provide a means for \emph{choosing}
$k$.

Let $\mathbf{K} = \{k_1, k_2, \dots , k_L\}$ be a vector of possible $k$-values for the
next measurements, and let $\mathbf{T} = \{t_1, t_2, \dots , t_L\}$ be the vector of
corresponding times; i.e. $t_n$ is the time required to perform an experiment with
$k = k_n$. The methods for choosing $k$ take the vectors $\mathbf{K}$ and $\mathbf{T}$ as
input. The value of the expected gain used by these methods is the maximum over possible
values of $\alpha \in [0, \pi]$.

\subsubsection{Multi-step gain method}
The first method we study for the adaptive choice of $k$ is based on the following idea:
if we compute the expected gain for more than one measurement we can compare the expected
knowledge gain for different sequences of measurements that take the \emph{same total
  time}. Comparing many possible $k$-values in this way will generally lead to an
expensive optimisation since there can be many measurement sequences that take the same
total time, and because the complexity of computing the expected knowledge gain grows
exponentially with the number of measurements (also see the Supplemental Material, section
\ref{sec:sup-multi-step}). Therefore we devise a technique which allows us to compare only
a few $k$-values at a time.

The following example illustrates how this method proceeds. Suppose the amount of
resources per $k$-value is proportional to $k$ ($\mathbf{T} \propto \mathbf{K}$) and we
restrict the choice of $k$ to the set $\mathbf{K} = \{2^{n-1}, n = 1, 2, \ldots, L\}$. We
can then compute the expected knowledge gain for performing two measurements with $k = 1$
versus the gain for performing one measurement with $k = 2$. If the former is greater, we
perform an experiment with $k = 1$. Otherwise, we can compute the expected knowledge gain
for performing two measurements with $k = 2$ versus that for performing one measurement
with $k = 4$. If the former is greater, we perform an experiment with $k = 2$, and so
on. Note that calculating the expected gain for multiple measurements requires optimising
one control phase $\alpha$ for each measurement. This optimisation is performed
sequentially starting with the first measurement. A detailed description is given in the
Supplemental Material, section \ref{sec:sup-multi-step}.

In this example we see that we need only compute the expected knowledge gain for one or
two measurements and for only a few possible sequences of measurements. We expect this to
converge to a locally optimal choice of $k$. This method for choosing $k$ can be
generalised to other possible input vectors $\mathbf{K}$ and $\mathbf{T}$ using an
algorithm which is given in the Supplemental Material, section
\ref{sec:sup-multi-step}. It works with vectors $\mathbf{K}$ and $\mathbf{T}$ sorted in
increasing order, under the assumption that $k_j > k_i$ implies $t_j \ge t_i$, and
restricted to $t_i/t_{i-1} < 32/7$. In this general algorithm we restrict the computation
of expected knowledge gain to at most 5 measurements to prevent the computation from
becoming too expensive. To allow for general inputs $\mathbf{K}$ and $\mathbf{T}$ under
this restriction, we in general compare sequences that require only \emph{approximately}
the same time. The detailed method is described by algorithm \ref{alg:multi-step-method}
in the Supplemental Material, section \ref{sec:sup-multi-step}.

\subsubsection{Gain rate method}
The second method we study for choosing $k$ uses the \emph{rate} of expected knowledge
gain. A similar idea has been used recently to estimate decoherence timescales in a qubit
\cite{2022arshad}. We compute a vector of expected knowledge gain
$\mathbf{G} = \{g_1, g_2, \dots , g_L\}$ where $g_n$ corresponds to the expected knowledge
gain for performing a single measurement with $k = k_n$, computed using either
\eqref{eq:sharpness-gain} or \eqref{eq:entropy-gain} depending on the choice of gain
function. We then calculate the vector of expected \emph{rate} of knowledge gain
$\mathbf{R} = \mathbf{G}/\mathbf{T}$ (elementwise division), and choose the $k = k_{n^*}$
with $n^* = \arg\max_n \mathbf{R}$. The value of $n^*$ may be found by performing a brute
force search over $k$-values. Since we use a computation that runs in series, the brute
force search can take too long for many practical implementations, and we reduce the
computation by performing a Fibonacci search \cite{1966avriel} to find a good value of
$k$. Since the computation of any element of $\mathbf{R}$ can be performed independently,
a brute force search could be performed by computing the gain rate for each $k$-value in
parallel. Since the brute force search is expected to lead to slightly better performance
than the Fibonacci search we use here, it could be considered for future
implementations. Overall, the gain rate method has the advantage of being simpler and
computationally cheaper than the multi-step method.

\section{\label{sec:num-rep}Numerical representations}

The representation of phase knowledge using the Fourier series \eqref{eq:density} means
that the number of non-zero coefficients $c_n$ to keep track of increases with our
knowledge of the phase. If the number of non-zero coefficients at step $s$ is
$\Gamma$, the computation for the Bayesian update \eqref{eq:update-rule} requires
calculating $\Gamma + k$ new coefficients. Considering the case that we start with a
uniform probability density, we initially have $c_0 = 1,\, c_n = 0,\, \forall n \neq 0$ in
\eqref{eq:density}. After performing some measurements with $N$ applications of $U$ in
total, the last update will require calculating $\Gamma = N$ non-zero coefficients.

If we are in a situation which results in Heisenberg scaling, then
$N \propto 1/\Delta\hat{\phi}$. If the scaling of $\Delta\hat{\phi}$ is slower than
Heisenberg scaling, then it follows from \eqref{eq:update-rule} that the number of
coefficients $\Gamma$ after reaching a given value of $\Delta\hat{\phi}$ will be greater
than for Heisenberg scaling. However, if we assume a probability density $p(\phi)$ which
on average preserves its functional form (i.e. the overall shape stays the same and only
the width changes) as estimation proceeds and $\Delta\hat{\phi}$ decreases, then from
standard Fourier analysis $\Gamma$ should scale as $1/\Delta\hat{\phi}$. This follows from
the fact that we expect the square root of the Holevo variance \emph{of the Bayesian
  probability density}, $\sqrt{V[p(\phi)]}$, to scale in the same way as
$\Delta\hat{\phi}$ with $N$. This suggests that when we converge more slowly than
Heisenberg scaling, the description of $p(\phi)$ obtained from \eqref{eq:update-rule} uses
disproportionately many non-zero coefficients and becomes inefficient. For this case, we
should seek a good representation of $p(\phi)$ with fewer coefficients to minimise
computational overhead. In this sense we can expect that it should be possible for the
number of coefficients to be inversely proportional to the uncertainty,
$\Gamma \propto 1/\Delta\hat{\phi}$, even if we do not have Heisenberg scaling.

The computation for adaptively choosing $\alpha$ and $k$ scales more favourably with
$\Gamma$ than the Bayesian update. Computing the expected sharpness gain
\eqref{eq:sharpness-gain} does not scale with the total number of coefficients $\Gamma$,
while the expected entropy gain \eqref{eq:entropy-gain} scales as $\Gamma/k$. Since we
expect the optimal value of $k$ to increase with $N$ (in the noise-free case), there may
exist algorithms, offering a complexity independent of $N$, for choosing $k$ such that the
expected entropy gain typically is maximised.

The practicality of using TAPE depends on the possibility of computing the Bayesian update
and choice of control parameters ($\alpha$ and $k$) in a time shorter or equal to the time
it takes to run the next measurement. Since the update computation time scales linearly
with $\Gamma$, it will become longer than the measurement time for sufficiently large
$\Gamma$. In the asymptotic limit when $N\rightarrow\infty$, the average measurement time,
proportional to $k$ for sequential procedures, will also scale linearly with $N$ if we
have Heisenberg scaling. If the slope of the linear increase in computation time is less
than that of the measurement time, the computation will remain shorter than the
measurement time on average for all $N$. However, this may not be the case for many
experiments where TAPE could otherwise be useful, for example if decoherence prevents
large $k$-value measurements from giving phase information, or if TAPE is used for
parallel procedures. In these cases it is beneficial to find methods that reduce the time
of computation. One approach is to change representation from a truncated Fourier series
to another function such as a Gaussian which is described by a number of parameters that
is independent of $N$ \cite{2021vandenberg}. Here we propose an alternative method that
allows the truncated Fourier series representation to be used for arbitrarily large $N$
while ensuring that the number of coefficients needed in the series scales much slower
than $1/\Delta\hat{\phi}$. This way all the procedures of sections \ref{sec:bayes-est} and
\ref{sec:adaptive-proc} can still be used for large $N$ with minimal changes, and the user
is also free to adjust the generality of the description of $p(\phi)$ versus computational
overhead as suits their purposes.

Given a probability density $p(\phi)$, we define a \emph{contraction} of $p(\phi)$ as
$(M, \phi_0, q(\theta))$, such that $p(\phi_0 + \theta/M) \approx q(\theta)$ for
$0 \le \theta < 2\pi$ and $p(\phi) \approx 0$ otherwise. We refer to $M \in \mathbb{N}^+$
as the magnification, to $\phi_0 \in [0, 2\pi)$ as the offset, and we represent
$q(\theta)$ as a truncated Fourier series. Thus, by representing the density $p(\phi)$ as
a contraction we use the truncated Fourier series representation for
$\phi_0 \le \phi < \phi_0 + 2\pi/M$ and assume that $p(\phi) = 0$ for values of $\phi$
outside this interval. We are effectively ``zooming in'' on a region around the expected
value of $\phi$ and assuming that the probability of $\phi$ outside this region is
negligible. Given a contraction $(M, \phi_0, q(\theta))$ where $q(\theta)$ has
coefficients $c_n,\, \forall n = -\Gamma, \dots, \Gamma$ we can calculate a new
contraction $(M', \phi_0', q'(\theta'))$, $M'$ a multiple of $M$ with
$m = M'/M \in \mathbb{N}^+$ as
\begin{align}
  \phi_0' &= \phi_0 + \hat{\theta}/M - \pi/M' \\
  c'_n &= \Exp{i(\pi-\arg(c_{-m}))}c_{m\times n},\,
         \forall n = -\Gamma/m, \dots, \Gamma/m \, ,
\end{align}
where $\hat{\theta} = \arg(c_{-1})$.

By using the truncated Fourier series on a reduced interval of size $2\pi/M$, a
contraction $(M, \phi_0, q(\theta))$ of $p(\phi)$ (with $\Gamma$ coefficients) uses only
$\Gamma/M$ coefficients. The assumption that $p(\phi) = 0$ outside the interval
${[\phi_0, \phi_0 + 2\pi/M)}$ is often false with finite probability. However by
performing more measurements this probability can always be reduced arbitrarily while
increasing the number of coefficients $\Gamma$. Once this probability is small enough for
the given application we can make contractions to ensure that the number of coefficients
no longer exceeds $\Gamma$. Thus, with this method we must balance the probability that a
contraction fails to represent the estimated phase against the maximum computation time
per measurement.

In the Supplemental Material, section \ref{sec:sup-cnt-cpx}, we show that under the
assumption of a Gaussian probability density $p(\phi)$, using contractions reduces the
number of Fourier coefficients required by $\mathcal{O}(\sqrt{\log(\log(\Gamma))})$, while
keeping the probability that $p(\phi) \neq 0$ outside the interval
$[\phi_0, \phi_0 + 2\pi/M)$ below a constant value (which can be chosen to be arbitrarily
small). In general, the computational complexity will depend on the particular functional
form of $p(\phi)$, but the Gaussian case demonstrates that the method can sometimes
greatly reduce computation. If a much poorer complexity is found when faster computation
is needed, it could be helpful to modify the adaptive strategy for choosing $\alpha$ and
$k$ such that $p(\phi)$ becomes closer to a Gaussian before each contraction is performed.

The expected sharpness or entropy gain relations \eqref{eq:sharpness-gain} and
\eqref{eq:entropy-gain} can be used for the contracted Fourier series representation
$q(\theta)$ with the replacement $\phi \rightarrow \theta$, although the values of the
expected sharpness gain in that case will not correspond to those for the full density
$p(\phi)$ \footnote{Although the differential entropy is not invariant when changing the
  scale, the differential entropy \emph{gain} is.}. However the control parameters that
achieve the maximum gain for the contraction can still be used to determine the values to
use for the next measurement. Denoting the optimal control phase thus obtained as $\beta$
and the optimal number of applications of $U$ as $j$, then the corresponding values to use
for the un-contracted density $p(\phi)$ can be calculated as $k = M j$,
$\alpha = \beta + k\phi_0$. In this way we can use contractions to reduce the time not
only for the update computation but also for the search of optimal parameters $\alpha$ and
$k$ using the expected entropy gain \footnote{this also works when using sharpness gain,
  but there is no speedup}.

\section{Simulation results}

In order to characterise the performance of the methods considered above we study the
results of numerical simulations. We focus on estimation starting from a uniform prior
(initially no information about the phase), though we note that our methods are also
well-suited to any prior that can be well-represented with a Fourier series and the
contractions described in section \ref{sec:num-rep}, given available computational
resources and particular timing requirements of an experiment. We first consider the case
of noise-free quantum metrology. In this context the time of the measurement is assumed to
be proportional to the number of applications $k$ of the unitary evolution $U$; in the
following we set the time of $U$ to 1 so that the time of a single measurement is equal to
$k$. In this case, the number of resources $N$ in the SQL,
$\Delta\hat{\phi} > 1/\sqrt{N}$, and HL, $\Delta\hat{\phi} > \pi/N$, is the total number
of applications of $U$ in the estimation procedure, and is equal to the total estimation
time.

\begin{figure}[h]
  \centering
  \includegraphics[width=0.48\textwidth]{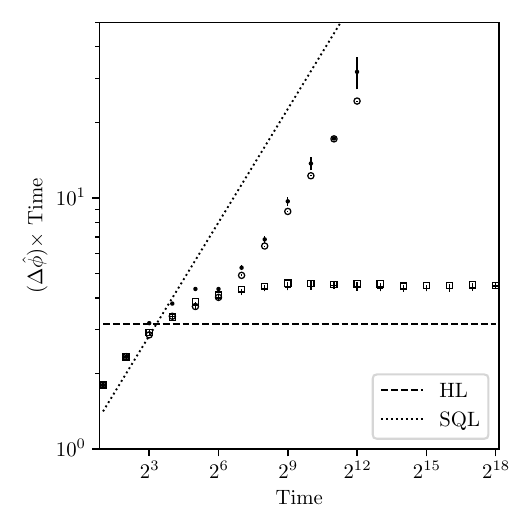}
  \caption{Metrology scaling without noise. The time of all experiments in an estimation
    sequence is equal to the total number of applications of the unitary $U$. The phase
    uncertainties $\Delta\hat{\phi}$ are calculated from $6\times 10^4$ ($3\times 10^6$)
    realisations of estimation for the multi-step (gain rate) methods with system phase
    $\phi$ chosen uniformly at random. Error bars (typically too small to see) are
    calculated by error propagation from sample standard deviations of the estimation
    errors. Legend: multi-step method maximising expected sharpness gain ($\circ$),
    multi-step method maximising expected entropy gain ($\bullet$), gain rate method
    maximising expected sharpness gain ($+$), hybrid gain rate method ($\square$).}
  \label{fig:metrology-scaling}
\end{figure}

In figure \ref{fig:metrology-scaling} we plot the uncertainty in the phase estimates
resulting from repeating the estimation procedure with randomly chosen values for the
system phase $\phi$ \footnote{The performance of estimation may be dependent on the system
  phase when starting from a uniform prior if the control phase for the first measurement
  is fixed (e.g. $\alpha = 0$). By choosing a random control phase $\alpha \in [0, \pi]$
  for the first measurement, the performance is the same for any value of the system
  phase. Thus, our results can also be interpreted as the performance for any particular
  value of the system phase assuming a random control phase is used in the first
  measurement.}. More specifically, we plot the uncertainty multiplied by total estimation
time for which any phase estimation reaching HS results in a constant value independent of
estimation time; for the HL, the constant value is $\pi$. The uncertainty in the phase
estimates $\Delta\hat{\phi}$ is calculated according to equation \eqref{eq:Holevo-var},
where the sharpness is calculated as $S(\hat{\phi}) = \expect{\cos(\hat{\phi} - \phi)}$
for a true system phase $\phi$, and the expectation value is approximated as an average
over simulations. This choice can only lead to larger uncertainty values than the usual
sharpness, $|\expect{\Exp{i(\hat{\phi} - \phi)}}|$, since
$\expect{\cos(\hat{\phi} - \phi)} = \mathfrak{Re}\{\expect{\Exp{i(\hat{\phi} -
    \phi)}}\}$. For the results in figure \ref{fig:metrology-scaling} we find
$\mathfrak{Im}\{\expect{\Exp{i(\hat{\phi} - \phi)}}\} \ll
\mathfrak{Re}\{\expect{\Exp{i(\hat{\phi} - \phi)}}\}$, which shows that the estimator is
unbiased and either choice of sharpness measure leads to essentially the same values for
the uncertainty. Nevertheless, in all simulation results presented in this manuscript we
have calculated the phase uncertainty $\Delta \hat{\phi}$ using
$S(\hat{\phi}) = \expect{\cos(\hat{\phi} - \phi)}$.

For the multi-step (gain rate) method, each plotted point is
$(\Delta\hat{\phi})\times (\text{total estimation time})$ calculated from $6\times 10^4$
($3\times 10^6$) realisations of estimation. We compare four adaptive methods for the
choice of control parameters $\alpha$ and $k$. For the multi-step method we plot
separately the results of maximising either the expected sharpness gain
\eqref{eq:sharpness-gain} or the expected entropy gain \eqref{eq:entropy-gain} for total
estimation times $2^m,\, m = 1, \dots , 12$. For the gain rate method we plot the results
of maximising the expected sharpness gain \eqref{eq:sharpness-gain} to determine the
choices of $\alpha$ and $k$, but we find that maximising the expected entropy gain
\eqref{eq:entropy-gain} performs worse (not shown), which is due to the properties of
entropy mentioned in section \ref{sec:know-gain}. We also note that when maximising the
expected entropy gain only, we tend to obtain very accurate estimates with a small number
of outliers that significantly increase the phase uncertainty.

Instead of studying further the gain rate method maximising only expected entropy gain, we
plot the result of a hybrid strategy in which the choices of $\alpha$ and $k$ are
determined using the expected entropy gain for at most the first half of the total
estimation time, and the expected sharpness gain is used for the remaining time. This
approach is motivated by the fact that although maximising ones information about the
phase is desired, obtaining a precise estimate for the phase additionally requires a
narrow probability density $p_s(\phi)$. Since multi-peak density functions can have the
same entropy as single-peak functions with a broader peak, maximising the entropy doesn't
necessarily lead to a narrow density. We choose the hybrid method to study the performance
of strategies that maximise the information gain, while still ensuring a narrow density
$p_s(\phi)$ that is required for accurate phase estimation.

As shown by Guti\'{e}rrez-Rubio et al. \cite{2020gutierrez-rubio}, maximising the entropy
gain leads to the maximum likelihood of estimating the correct system phase. Another way
to understand the drawback of maximising the entropy gain only, is that maximising the
likelihood of estimating the system phase $\phi$ does not generally minimise
$\Delta\hat{\phi}$, which depends on the entire distribution of phase estimates, call it
$q(\hat{\phi})$, rather than the value $q(\hat{\phi} = \phi)$ at a single point. Note that
if one could maximise expected gain over an entire measurement sequence used for
estimation, maximising the expected sharpness gain should minimise $\Delta\hat{\phi}$
\cite{2009berry}. Since we perform the optimisation only for the next measurement, it
becomes interesting to study the hybrid method.

The two plotted methods based on gain rate are for total estimation times
$2^m,\, m = 1, \dots , 18$, and all methods make use of contractions (section
\ref{sec:num-rep}) to reduce computation times; if the square root of the Holevo variance
of the phase density, $\sqrt{V[p(\phi)]}$ is less than $\frac{\pi}{2^{13} M}$, where $M$
is the current magnification, then we perform at least one measurement using parameters
$\alpha, k$ that maximise the expected sharpness gain, and then we perform a contraction
with $m=2$. The additional measurement to maximise the expected sharpness gain is used in
the hybrid method, and multi-step method maximising entropy to reduce the probability of
the system phase being outside the reduced interval over which the Fourier series
representation of the phase knowledge is used in the contraction. Since $k$-values are
chosen adaptively, a sequence of optimal choices does not generally lead to the total
times we choose to plot in figure \ref{fig:metrology-scaling}. For the simulations plotted
we therefore always restrict the optimisation over $k$ to values less than or equal to the
total \emph{remaining} time. For this reason we have performed simulations for each
plotted point independently of the other points for different total times.

The results in figure \ref{fig:metrology-scaling} show that all methods perform better
than the SQL. The gain rate method using sharpness gain or the hybrid method perform best,
and we fit the results for total estimation times from $2^{12}$ to $2^{18}$ with a
functional form $\Delta\hat{\phi} = A \pi N^{-\gamma}$. The values of $A$ and $\gamma$
obtained from the fits are summarised in table \ref{tab:metrology-scaling}. The hybrid
method clearly reaches HS ($\gamma = 1$), while the method maximising sharpness only is
extremely close to HS, and both methods are close to the theoretically lowest achievable
uncertainty for the values simulated. In particular, the uncertainty for a total time of
$2^{18}$ (i.e. the longest time simulated) is a factor of $1.422 \pm 0.015$ from the HL
when maximising sharpness gain only. For the same total time the hybrid method reaches an
uncertainty within a factor of $1.429 \pm 0.011$ from the HL.

\begin{table}[H]
  \begin{ruledtabular}
    \begin{tabular}[c]{lcc}
      Method & $A$ & $\gamma$ \\
      sharpening & $1.371 \pm 0.028$ & $0.998 \pm 0.002$ \\
      hybrid & $1.430 \pm 0.036$ & $1.0 \pm 0.002$ \\
    \end{tabular}
  \end{ruledtabular}
  \caption{Values of the fitted parameters $A$ and $\gamma$ using the fit function
    $\Delta\hat{\phi} = A \pi N^{-\gamma}$ and the uncertainties obtained for estimation
    times from $2^{12}$ to $2^{18}$ for the two gain rate methods plotted in figure
    \ref{fig:metrology-scaling}.}
  \label{tab:metrology-scaling}
\end{table}

We attribute the poorer performance of the multi-step method to the fact that the compared
gains require only approximately equal time and to the fact that only a local optimum is
used (see Supplemental Material, section \ref{sec:sup-multi-step}). Although we expect the
performance could be improved by starting the search at a $k$-value dependent on the
Holevo variance of the prior (cheap to compute) and additionally using a ``search down''
method analogous to algorithm \ref{alg:search-up} (see Supplemental Material, section
\ref{sec:sup-multi-step}), we choose rather to focus on the gain rate method since it
demonstrates near-optimal performance in addition to being simpler and computationally
more efficient.

In the Supplemental Material, section \ref{sec:sup-k-sub}, we have also compared the
multi-step and gain rate methods in the case where we allow only the subset of $k$-values:
$k \in \{2^n\},\, n = 1,2,3,\dots$. In this case the gains compared by the multi-step
method require \emph{exactly} the same time, and the method performs similarly to the gain
rate method. However even in this situation the multi-step method at best performs
similarly to the hybrid method using the gain rate. For completeness, we have also
simulated the gain rate methods studied here when a brute force search for $k$ is used
rather than a Fibonacci search. In that case the hybrid method performs best reaching an
uncertainty within $1.394 \pm 0.010$ of the HL. A summary of simulation results for the
best performing methods we have studied is given in table \ref{tab:scaling} in the
Supplemental Material, section \ref{sec:sup-k-sub}.

\begin{figure}[h]
  \centering
  \includegraphics[width=0.48\textwidth]{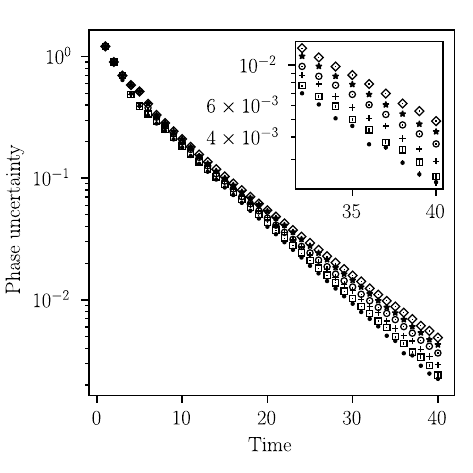}
  \caption{Scaling without noise when the experiment time is equal to the number of
    measurements (independent of how many applications of $U$ are used). Plotted are the
    phase uncertainties $\Delta\hat{\phi}$ calculated from $12 \times 10^6$ realisations
    of estimation with system phase $\phi$ chosen uniformly at random. Knowledge gains
    optimised with rate from metrology scaling ($\mathbf{T} \propto \mathbf{K}$): expected
    sharpness gain ($\diamond$), expected entropy gain ($\circ$), hybrid method
    ($\star$). Knowledge gains optimised with the correct rate
    ($\mathbf{T} = \{1,1,\dots ,1\}$): expected sharpness gain ($+$), expected entropy
    gain ($\bullet$), hybrid method ($\square$). Inset: closer view of results for 32 to
    40 measurements.}
  \label{fig:shot-scaling}
\end{figure}

In figure \ref{fig:shot-scaling} we consider the performance of the methods using the gain
rate when the measurement time is independent of $k$. This situation describes the limit
where the time required to perform the unitary evolution $U$ is negligible compared to
other times in the experiment such as state preparation and readout. In this case the
total time is equal to the number of measurements. We plot methods based on maximising
either sharpness gain or entropy gain as well as a hybrid method, as in figure
\ref{fig:metrology-scaling}. In addition we compare the case where the gain maximisation
is based on the correct rate in this context, i.e.  $\mathbf{R} = \mathbf{G}/\mathbf{T}$
where $\mathbf{T} = \{1,1,\dots ,1\}$ (so $\mathbf{R} = \mathbf{G}$), with the usual case
in metrology: $\mathbf{T} \propto \mathbf{K}$. We plot the mean error resulting from
$12\times 10^6$ realisations of estimation from 1 to 40 measurements.

The results in figure \ref{fig:shot-scaling} show that the strategies using the correct
rate ($\mathbf{R} = \mathbf{G}$) outperform those optimised for the usual metrology
setting with $\mathbf{T} \propto \mathbf{K}$. In this case we find that the method that
maximises the expected entropy gain performs best while that maximising expected sharpness
gives the lowest accuracy in phase estimation; the hybrid method has a performance in
between the two. We fit the results from $t = 10$ to $40$ measurements with an exponential
decay $\Delta\hat{\phi} = A\exp{(-\kappa t)}$. The values of the fitted parameters $A$ and
$\kappa$ for each of the plotted curves in figure \ref{fig:shot-scaling} are listed in
table \ref{tab:shot-scaling}.

\begin{table}[H]
  \begin{ruledtabular}
    \begin{tabular}[c]{lcc}


      Method & $A$ & $\kappa$ \\
      sharpening (M) & $0.7659 \pm 0.0109$ & $0.13164 \pm 0.00102$ \\
      hybrid (M) & $0.7903 \pm 0.0139$ & $0.13536 \pm 0.00127$ \\
      entropy (M) & $0.7984 \pm 0.0119$ & $0.14074 \pm 0.00103$ \\
      sharpening & $0.8015 \pm 0.0031$ & $0.14149 \pm 0.00028$ \\
      hybrid & $0.7739 \pm 0.0018$ & $0.14424 \pm 0.00017$ \\
      entropy & $0.7829 \pm 0.0027$ & $0.14856 \pm 0.00026$ \\
    \end{tabular}
  \end{ruledtabular}
  \caption{Values of the fitted parameters $A$ and $\kappa$ for the fit function
    $\Delta\hat{\phi} = A\exp{(-\kappa t)}$ for each of the methods plotted in figure
    \ref{fig:shot-scaling}. (M) indicates the rate used for optimisation is the one
    usually considered in quantum metrology, $\mathbf{T} \propto \mathbf{K}$.}
  \label{tab:shot-scaling}
\end{table}

Although the performance is better when the correct rate is used, the improvement is not
dramatic. This is perhaps not surprising since the quantum phase estimation algorithm
(QPEA) \cite{1998cleve, 2010nielsen} is known to be optimal even when the number of
applications of $U$ is not the relevant resource (indeed this is also the case in the
setting of quantum computation) \cite{2007vandam, 2009wiseman}. Methods based on QPEA also
lead to Heisenberg scaling in the context of quantum metrology \cite{2009berry}; that is,
estimation procedures using similar allocation of $U$ per measurement have been shown to
be optimal in both settings. Nevertheless, we see that in the case of the single-step
optimisation we perform for sequential strategies we can obtain a slight improvement by
using the appropriate rate in the optimisation. In this setting we attribute the better
performance when maximising the information gain rather than the sharpness gain to the
fact that higher $k$-values are cheap compared to the metrology setting, allowing
strategies that are better quantified by the richer nature of the entropy.

For the remaining simulations we return to the metrology setting where
$\mathbf{T} \propto \mathbf{K}$ since in this context we can compare performance with
known theoretical bounds. In particular, we study the performance in the presence of
noise.

\begin{figure}[h]
  \centering
  \includegraphics[width=0.48\textwidth]{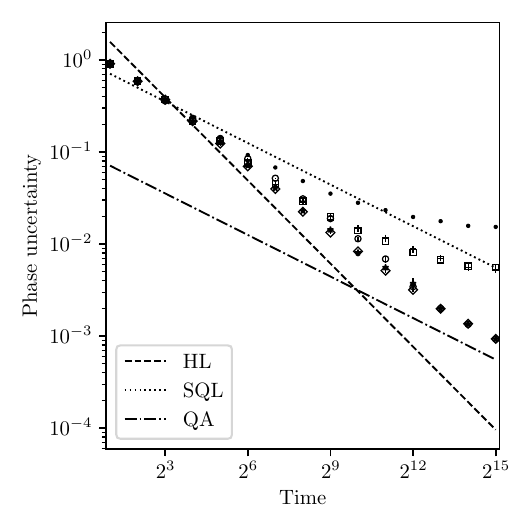}
  \caption{Metrology scaling in the presence of decoherence. Plotted are the phase
    uncertainties $\Delta\hat{\phi}$ calculated from $32\times 10^4$ realisations of
    estimation with system phase $\phi$ chosen uniformly at random. HL: Heisenberg
    limit. SQL: standard quantum limit. QA: lower bound on achievable quantum advantage
    given by $c/\sqrt{N}$, where $c = \sqrt{1-\eta^2}/\eta$, $\eta = 0.995$, and $N$ is
    the total time. Estimation with, i.e. $\zeta_k = \Exp{-(1-\eta)k},\, \forall k$
    (without, i.e.  $\zeta_k = 1,\, \forall k$), decoherence included in the estimator
    model: sharpness: $\diamond$ ($+$), entropy: $\circ$ ($\bullet$), hybrid: $\star$
    ($\square$).  $\diamond$ and $\star$ are mostly overlapping since the corresponding
    methods perform similarly.}
  \label{fig:decoherence-scaling}
\end{figure}

In figure \ref{fig:decoherence-scaling} we model a system with dephasing. We assume
perfectly prepared $\ket{+} \equiv (\ket{0} + \ket{1})/\sqrt{2}$ states, and
$U = \Exp{-iZ\phi/2}$, $Z$ the Pauli-z operator. After each application of $U$ we
additionally apply a dephasing channel described by the Kraus operators
$K_0 = \sqrt{\frac{1+\eta}{2}}\openone$, $K_1 = \sqrt{\frac{1-\eta}{2}}Z$. At the end we
perform the measurement in the x-basis. In this case it is possible to have Heisenberg
scaling only initially while increasing $k$ is beneficial, but as $k$ is increased
dephasing eventually reduces the information available by the measurement and we are
restricted to $1/\sqrt{N}$ scaling. The ultimate bound on precision can then be expressed
as $c/\sqrt{N}$, where $c$ is a prefactor depending on the noise channel. In the case of
dephasing this prefactor has been shown to be equal to at least $\sqrt{1-\eta^2}/\eta$
(but it is not known if this bound is tight) \cite{2012demkowicz-dobrzanski}. In figure
\ref{fig:decoherence-scaling} we consider the case where $\eta = 0.995$; the bound
$c/\sqrt{N}$ is also plotted. As before we simulate the gain rate method for sharpness
gain, entropy gain, and the hybrid case described above. We compare these three methods
when no decoherence is accounted for in the estimator model ($\zeta_k = 1$ in equation
\eqref{eq:meas-prob}) and when the decoherence rate is already known
($\zeta_k = \Exp{-(1-\eta)k}$). We plot the phase uncertainties calculated from
$32\times 10^4$ realisations of the estimation starting from a uniform prior and system
phase chosen uniformly at random.

When decoherence is not accounted for in the estimator model, maximising the entropy leads
to the poorest performance since this chooses larger $k$-values for which the system
decoheres, leading to reduced information in the measurement. For all methods that do not
account for decoherence in the model, estimation works initially, but as $k$ is increased
decoherence eventually introduces large random errors in the estimates. When decoherence
is accounted for in the model we find that all methods perform well. When sufficiently
many resources are used the method that maximises the rate of entropy gain performs
similarly to the other methods. This can be explained by the fact that decoherence
eventually limits the $k$-values chosen by the strategy and the density $p_s(\phi)$ is far
less likely to have multiple peaks. We find that when the decoherence is accounted for,
the performance after a total estimation time of $2^{15}$ is within $1.6715 \pm 0.0021$ of
the theoretical bound when maximising the expected entropy gain, $1.6807 \pm 0.0063$ when
maximising the expected sharpness gain, and $1.6734 \pm 0.0021$ for the hybrid method. Our
methods could also be combined with an optimised estimation of decoherence timescales as
proposed by Arshad et al. \cite{2022arshad}.

\begin{figure}[h]
  \centering
  \includegraphics[width=0.48\textwidth]{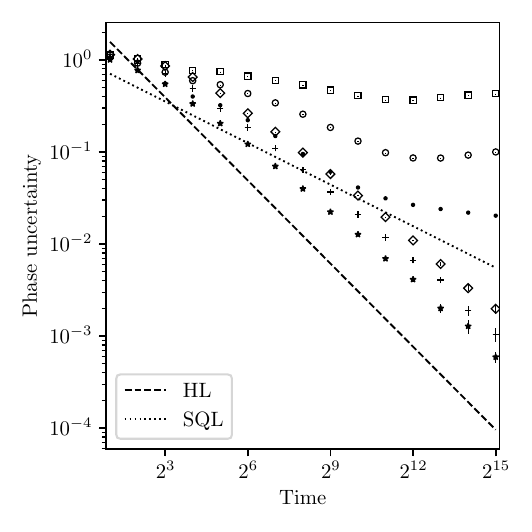}
  \caption{Metrology scaling with bit-flips and spontaneous emission. Each point is the
    phase uncertainty $\Delta\hat{\phi}$ calculated from $2\times 10^6$ realisations of
    estimation with system phase $\phi$ chosen uniformly at random. The hybrid method
    described in the main text is used for all simulations plotted here.  \emph{Without}
    noise included in the model of the estimator, i.e.
    $\lambda_k = \zeta_k = 1,\, \forall k$: $p = 0.3$ ($\square$), $p = 0.2$ ($\circ$),
    $p = 0.1$ ($\bullet$). \emph{With} noise included in the estimator model, i.e.
    $\lambda_k = \zeta_k = (1-p),\, \forall k$: $p = 0.3$ ($\diamond$), $p=0.2$ ($+$),
    $p=0.1$ ($\star$).}
  \label{fig:robustness}
\end{figure}

The results in figure \ref{fig:decoherence-scaling} demonstrate some robustness of the
estimation methods to errors not accounted for in the model. In figure
\ref{fig:robustness} we present the results of a last set of simulations to examine
further the robustness of the methods. In this case we assume perfectly prepared $\ket{0}$
states and $U = \Exp{iY\phi/2}$. After the $k$ applications of $U$ we apply a bit-flip
channel with Kraus operators $K_0 = \sqrt{1-p_b}\openone$, $K_1 = \sqrt{p_b}X$, $X,Y$
Pauli-x,y, followed by spontaneous emission with Kraus operators
\begin{align*}
  K_0 &=
        \begin{pmatrix}
          1 &0 \\
          0 &\sqrt{1-p_s}
        \end{pmatrix},\quad K_1 =
              \begin{pmatrix}
                0 &\sqrt{p_s} \\
                0 &0
              \end{pmatrix} \, .
\end{align*}
We choose to set $p_b = p/2$, $p_s = p$ and simulate the hybrid gain rate method with
$p = 0.1, 0.2, 0.3$. For each value of $p$ we simulate estimation using an estimator model
with ($\lambda_k = \zeta_k = (1-p),\, \forall k$) and without
($\lambda_k = \zeta_k = 1,\, \forall k$) errors included.

We see that when errors are included in the model the phase uncertainty is significantly
larger than in the noise-free case, but we still retain better than classical scaling for
longer estimation times. We fit the phase uncertainty with a functional form
$\Delta \hat{\phi} = A \pi N^{-\gamma}$, with fit parameters $A$ and $\gamma$. By fitting
subsets of points we see that $\gamma$ tends to increase for longer times, but we would
need to simulate further to see if HS is eventually reached. When errors are not included
in the model, only the $p = 0.1$ case performs similarly to the SQL initially, while for
larger errors the performance drops significantly. We see also that for longer estimation
times the lack of errors in the model can prevent further reduction of the phase
uncertainty. This suggests that it may be better to restrict to lower $k$-values in the
presence of large unknown errors. Overall, these results demonstrate robustness of the
method to large errors indicating that TAPE can also be useful for calibration of quantum
systems, e.g. for quantum computing, as considered in ref. \cite{2021russokirby}.

In the noise-free case we find average computation times are $\sim 1\, \mathrm{ms}$ for
the computation required for a single shot using a standard PC (CPU model: Intel(R)
Core(TM) i7-8565U CPU @ 1.80GHz) without any parallel computation. When noise is included
in the model we find that the maximisation of expected knowledge gain can take a time up
to $\sim 3\, \mathrm{ms}$ for the entropy gain, and $\sim 1\, \mathrm{ms}$ when maximising
the expected sharpness gain. This makes the method suitable for trapped ion or neutral
atom based quantum computing systems where the time of a single shot is typically a few
milliseconds. Since the computation of expected knowledge gain for each $k$-value could be
performed in parallel it may be possible to obtain sufficient speedups for some
applications in superconducting circuits or NV centres. Further speedups may also be
possible when maximising the expected sharpness gain if the optimal value of $\alpha$
where determined analytically. An alternative approach to achieve faster computation with
the implementation we use here is to restrict measurements to use only a subset of
possible $k$-values; a further discussion can be found in the Supplemental Material,
section \ref{sec:sup-k-sub}.

For experimental systems with sufficiently low measurement rates, TAPE thus provides a
near-optimal and very flexible method for phase estimation. In the metrology setting
without noise, the best performance we know of for phase estimation using a classical
sequential strategy is the adaptive method of Higgins et al. that is inspired by QPEA from
quantum computing \cite{2007higgins, 2009wiseman}. Their adaptive method reaches a phase
uncertainty (as quantified by the square root of the Holevo variance of the estimates) a
factor of $1.56$ larger than the HL, while they have later devised a non-adaptive method,
also inspired by QPEA, that performs similarly, demonstrating uncertainty less than a
factor of $2.03$ larger the HL \cite{2009higgins, 2009berry}. The latter non-adaptive
method has also been shown to be a robust method for calibrating single-qubit gates for
quantum computation \cite{2021russokirby}. Non-classical strategies for phase estimation
reaching lower uncertainty estimates are known, as e.g. the method of Pezz\`e and Smerzi
\cite{2020pezze, 2021pezze} which provides phase estimates with uncertainty 1.27 larger
than the HL. However, classical sequential strategies are best suited to single-qubit gate
calibration. In the metrology setting we have found that TAPE reaches an uncertainty that
is a factor of $1.43$ larger than the HL using the hybrid method, demonstrating similar or
better performance to known methods. The results presented in figures
\ref{fig:decoherence-scaling} and \ref{fig:robustness} also show significant robustness to
errors, demonstrating that TAPE can be a good method for calibrating single-qubit
operations.

In contrast to algorithms inspired by QPEA where a predetermined number of measurements is
performed for values of $k \in 2^n,\, n = 1, 2, 3, \dots$, TAPE chooses the \emph{time} of
phase evolution (i.e. the $k$-value) adaptively. For the non-adaptive method that is used
for robust phase estimation \cite{2021russokirby, 2009higgins}, the number of measurements
for each $k$-value is optimised beforehand for a given total number of phase applications
$N = \sum_s k_s$ assuming no prior knowledge of the phase. If noise is present the
optimisation needs to be modified as shown in \cite{2020belliardo}. But as noted therein,
the noise models they consider ``are to be thought more as toy models'' that ``capture
some of the key features of those scenarios''. Depending on the type of noise present in a
given experiment a further analysis and optimisation of the number of measurements to
perform with each $k$-value will be required to minimise the phase uncertainty. By using
the general form for the measurement probabilities \eqref{eq:meas-prob}, TAPE allows for
the description of a wide range of noise to be included in the model of the estimator and
directly provides near-optimal phase estimation procedures by accounting for the modelled
noise in the optimisation for the control parameters $\alpha$ and $k$.

In addition, TAPE allows the exact experiment times to be easily included in the
optimisation for the adaptive choice of $k$-value. This is very convenient for experiments
where state preparation and readout times cannot be neglected in comparison to the time
required to apply the unknown phase. Using the QPEA inspired methods above, a further
optimisation of the number of measurements with each $k$-value would otherwise be needed
to minimise the phase uncertainty.

In the QPEA inspired methods above, the value of $k$ to use for a particular step in the
estimation sequence requires knowing how many measurements have been performed so far with
each $k$-value. In TAPE, once the model of the noise is specified (by setting the values
$\lambda_k,\, \zeta_k \, \forall k$), and the experiment times required for each $k$-value
are set, the values of the control parameters $\alpha$ and $k$ are determined using only
the current prior knowledge density $p(\phi)$. This makes it easy to apply TAPE in
situations where some prior knowledge may be available. As an example suppose we would
like to use phase estimation for calibrating single-qubit operations on a quantum computer
where internal parameters of the device can drift over time. One could use either TAPE or
a QPEA inspired method to initially estimate parameters. However, it would be easy to
include the drift rate in the model of the phase knowledge $p(\phi)$ as e.g. a broadening
of the probability density function over time. Then one could use TAPE to perform a
minimal number of measurements to keep track of the parameters needed for single-qubit
operations over time.

Other proposals for phase estimation such as \cite{2005mitchell, 2016wiebe, 2023smith},
and some discussed in \cite{2000berry} are more similar to TAPE in that they choose the
value of $k$ adaptively. However, they do not demonstrate better performance in terms of
uncertainty of phase estimates or flexibility in terms of using potentially available
prior information or accounting for experimental resources and imperfections.

\section{Conclusion}

Between the different forms of TAPE compared we find that choosing the control parameters
for the phase $\alpha$ and number of unknown phase applications $k$ based on the
\emph{rate} of knowledge gain gives near-optimal performance in several different
settings, while requiring computation times that make it accessible to many
experiments. In the context of noise-free quantum metrology we reach uncertainties in
phase estimates within $1.43$ of the HL using a hybrid method maximising the rate of
expected entropy and sharpness gains. In addition, we have found uncertainties within
$1.39$ of the HL using the hybrid gain rate method performing a brute-force search over
measurement setting $k$, rather than a Fibonacci search (Supplemental Material, section
\ref{sec:sup-k-sub}, table \ref{tab:scaling}). Performing the computations for each
$k$-value in parallel would allow this to be done in times comparable to those we find for
performing a computation in series with the Fibonacci search method, or even faster.

In a setting where experiment times are proportional to the number of measurements rather
than to the number of unknown phase applications $k$, we find that maximising the
information gain only, leads to the best performance, while the hybrid strategy performs
only slightly worse. The method is also able to find optimal strategies in the presence of
different types of noise, and demonstrates significant robustness to errors. Combined with
the fact that the optimisation can be easily tuned to the real times of experiments as a
function of $k$ and can be used with arbitrary prior information, TAPE thus provides an
extremely versatile phase estimation method that can directly give optimal performance in
a wide range of experimental settings.

\section*{Code availability}
The core implementation of TAPE used for all simulations in this work is available at:

{\small
  \url{https://bitbucket.org/brennann/qtape/src/main/}
}

\noindent The gain rate method is provided therein. The multi-step method was implemented
in \texttt{python} using the methods from the core implementation. The code used to
generate the figures and values simulated in this work are available from B.N. upon
reasonable request.

\section*{Author contributions}
Initial theory for phase estimation in the noise-free case for $k=1$ including derivation
of an expression for the expected entropy gain was done by A.V.L., and the resulting
adaptive method was implemented by V.N. for a trapped-ion experiment in the group of
J.P.H. to perform adaptive Ramsey measurements. B.N. had the idea to choose different
$k$-values adaptively using the same formalism, and worked out the theory in the general
case allowing for a range of noise. B.N. devised the gain rate and multi-step methods and
chose to study the sharpness as a measure of knowledge in addition to the entropy. The
manuscript was written by B.N. with input from all the authors.

\begin{acknowledgments}
  We acknowledge support from the Swiss National Science Foundation
  (SNF) under Grant No. 200020\_179147, and from Intelligence Advanced
  Research Projects Activity (IARPA), via the US Army Research Office
  grant W911NF-16-1-0070. B.N. thanks Ivan Rojkov for helpful feedback
  on the manuscript.
\end{acknowledgments}

\bibliography{../notes/tape_refs}

\part{}

\renewcommand{\thesection}{S.\Roman{section}}

\setcounter{figure}{0}
\renewcommand{\thefigure}{S.\arabic{figure}}

\setcounter{table}{0}
\renewcommand{\thetable}{S.\arabic{table}}

\setcounter{equation}{0}
\renewcommand{\theequation}{S.\arabic{equation}}

\setcounter{algorithm}{0}
\renewcommand{\thealgorithm}{S.\arabic{algorithm}}

\section{\label{sec:sup-k-sub} $k$-value subsets}

In order to simplify the optimisation procedure that is performed to choose $\alpha$ and
$k$ adaptively for each experiment it is interesting to consider procedures where we
restrict $k$-values to certain subsets. The main advantage is that the optimisation can be
performed in less time (without parallel computation) which could make TAPE accessible to
experimental systems where the time of a single shot is shorter. Here we focus on the
commonly studied subset containing only powers of two: $k \in \{2^n\},\, n = 1,2,3,\dots$.

In the particular case when $\mathbf{T} \propto \mathbf{K}$, which is usually studied in
quantum metrology the subset with only powers of two is also interesting to study for the
multi-step method; the results of section \ref{sec:sup-multi-step} show that the
multi-step method is limited by the fact that it converges to local maxima when optimising
the choice of $k$-value. This is due to the fact that the expected knowledge gains
compared are for sequences of experiments that only take \emph{approximately} equal
time. When $\mathbf{T} \propto \mathbf{K}$ and $k \in \{2^n\},\, n = 1,2,3,\dots$, the
gains compared by the multi-step method are for experiments that take \emph{exactly} the
same time, thereby avoiding the limitations of the multi-step method that occur in the
general case.
\begin{figure}[h]
  \centering
  \includegraphics[width=0.48\textwidth]{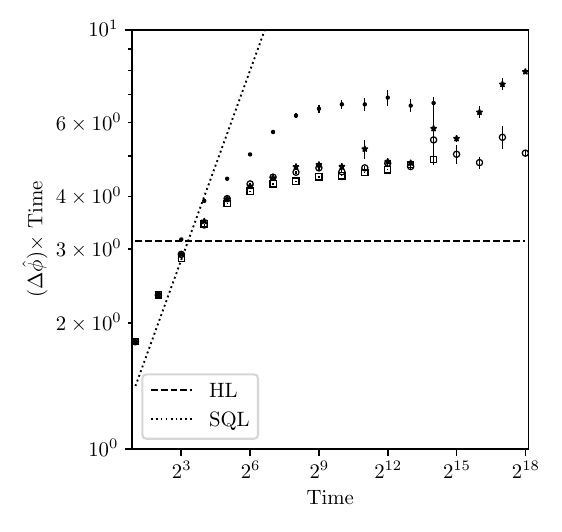}
  \caption{Metrology scaling without noise when only powers of $2$ are used as
    $k$-values. The time of all experiments in an estimation sequence is equal to the
    total number of applications of the unitary $U$. The phase uncertainties
    $\Delta\hat{\phi}$ are calculated from $10^6$ realisations of estimation with system
    phase $\phi$ chosen uniformly at random. For all methods plotted contraction is
    used. When $\sqrt{V[p(\phi)]}$, is less than $\frac{\pi}{2^{14} M}$ with current
    magnification $M$, then we perform at least one measurement using parameters
    $\alpha, k$ that maximise the expected sharpness gain, and then we perform a
    contraction with $m=2$. Expected entropy gain, multi-step method: $\bullet$, hybrid
    method (uses gain rate): $\circ$, expected sharpness gain, multi-step method (gain
    rate method): $\square$ ($\star$).}
  \label{fig:metrology-scaling-bin}
\end{figure}

We fit all results with a functional form $\Delta \hat{\phi} = A \pi N^{-\gamma}$. For the
multi-step method, we fit for total estimation times from $2^{12}$ to $2^{14}$, and for
the gain rate method from $2^{12}$ to $2^{18}$. The fit parameters as well as ratios to
the HL after $N = 2^{14}$ and $2^{18}$ are summarised in table \ref{tab:scaling}. Some
results from the main text for cases where all $k$-values are used are also included in
table \ref{tab:scaling} for comparison.

\begin{table*}[t]
  \begin{ruledtabular}
    \begin{tabular}[c]{lcccc}
      Method & $A$ & $\gamma$ & ratio to HL, $N = 2^{14}$ & ratio to HL, $N = 2^{18}$ \\
      hybrid, all $k$ (BFS) & --- & --- & --- & $1.394 \pm 0.010$ \\
      sharp., all $k$ (BFS) & --- & --- & --- & $1.400 \pm 0.013$ \\
      hybrid, all $k$ (FS) & $1.430 \pm 0.036$  & $1.0 \pm 0.002$ & $1.420 \pm 0.009$ & $1.429 \pm 0.011$ \\
      sharp., all $k$ (FS) & $1.371 \pm 0.028$ & $0.998 \pm 0.002$ & $1.390 \pm 0.010$ & $1.422 \pm 0.015$ \\
      hybrid, $k \in \{2^n\}$ (BFS) & $1.304 \pm 0.076$ & $0.983 \pm 0.006$ & $1.738 \pm 0.223$ & $1.617 \pm 0.026$ \\
      sharp., $k \in \{2^n\}$ (BFS) & $0.452 \pm 0.052$ & $0.862 \pm 0.010$ & $1.851 \pm 0.214$ & $2.528 \pm 0.019$ \\
      multi-step, entropy, $k \in \{2^n\}$ & $2.131 \pm 0.618$ & $1.0 \pm 0.018$ & $2.128 \pm 0.075$ & --- \\
      multi-step, sharp., $k \in \{2^n\}$ & $1.044 \pm 0.007$ & $0.958 \pm 0.001$ & $1.563 \pm 0.010$ & --- \\
    \end{tabular}
  \end{ruledtabular}
  \caption{Fit parameters for the methods plotted in figure
    \ref{fig:metrology-scaling-bin} using a functional form
    $\Delta \hat{\phi} = A \pi N^{-\gamma}$, and ratios to the HL after total estimation
    times of $N = 2^{14}$ and $2^{18}$. The results for the hybrid and maximum sharpness
    gain rate methods using contraction from figure \ref{fig:metrology-scaling} are also
    included for comparison. In addition we have put the results of gain rate methods when
    a brute force search is used; these particular simulations are results from
    $2 \times 10^6$ estimations with random system phase. BFS: brute force search. FS:
    Fibonacci search. All results are fitted for total estimation times $\ge 2^{12}$
    (except for BFS since we have only simulated the $N=2^{18}$ case).}
  \label{tab:scaling}
\end{table*}

\begin{table*}[t]
  \begin{ruledtabular}
    \begin{tabular}[c]{lcccc}
      Method & $t_{\mathrm{up}}$ & $t_{H}$ & $t_{S}$ & $t_{\mathrm{con}}$ \\
      hybrid, all $k$ (BFS) & $40\, (330)\, \mu s$ & $2.5\, (17)\, m s$ & $6\, (20)\, m s$ & $37\, (70)\, \mu s$ \\
      sharp., all $k$ (BFS) & $17\, (125)\, \mu s$ & --- & $3.2\, (20)\, m s$ & $27\, (50)\, \mu s$ \\
      hybrid, all $k$ (FS) & $42\, (330)\, \mu s$ & $800\, (4300)\, \mu s$ & $800\, (1700)\, \mu s$ & $38\, (70)\, \mu s$ \\
      sharp., all $k$ (FS) & $18\, (120)\, \mu s$ & --- & $450\, (1900)\, \mu s$ & $27\, (40)\, \mu s$ \\
      hybrid, $k \in \{2^n\}$ (BFS) & $90\, (700)\, \mu s$ & $250\, (1200)\, \mu s$ & $8\, (25) \mu s$ & $40\, (80)\, \mu s$ \\
      sharp., $k \in \{2^n\}$ (BFS) & $60\, (1000)\, \mu s$ & --- & $8\, (40)\, \mu s$ & $45 (100) \mu s$ \\
    \end{tabular}
  \end{ruledtabular}
  \caption{Benchmarks. A summary of approximate times required for computation with CPU
    model: Intel(R) Core(TM) i7-8565U CPU @ 1.80GHz. All values are rough estimates from
    running $\sim 20$ repetitions with total estimation time $N = 2^{16}$. In all
    simulations performed here we assume the experiment and modelled measurement
    probabilities are noise-free, and $\mathbf{T} \propto \mathbf{K}$. $t_{\mathrm{up}}$
    is the time required for the Bayesian update. $t_{H}$ is the time required to
    determine the optimal values of $\alpha$, and $k$ by maximising the expected entropy
    gain rate. $t_{S}$ is the time required for determining $\alpha$ and $k$ that maximise
    the expected sharpness gain rate. And $t_{\mathrm{con}}$ is the time required for
    contraction (when $\sqrt{V[p(\phi)]}$, is less than $\frac{\pi}{2^{12} M}$ then we
    perform at least one measurement using parameters $\alpha, k$ that maximise the
    expected sharpness gain, and then we perform a contraction with $m=2$). Values
    represent rough averages over all shots and runs of estimation. In parentheses are
    maximum values observed on a single shot over all runs of estimation. Methods are
    listed for brute force search (BFS) and Fibonacci search (FS) of the optimal
    $k$-value.}
  \label{tab:benchmarks}
\end{table*}

We see that gain rate methods optimising over all possible $k$-values reach HS, while
those using the subset $k \in \{2^n\},\, n = 1,2,3,\dots$ do not. However, the hybrid
method using the subset $k \in \{2^n\},\, n = 1,2,3,\dots$ is very close to HS and
performs only slightly worse than when all $k$-values are used. For methods allowing all
$k$-values we performed contractions when the Holevo variance of the phase density,
$V[p(\phi)]$ was less than $\frac{\pi}{2^{13} M}$, while when using only a subset of
$k$-values we had to use the condition $V[p(\phi)] < \frac{\pi}{2^{14} M}$ to sufficiently
suppress unwanted estimation errors. This suggests the probability distribution of phase
estimates can have larger tails when using only subsets of $k$-values. This in turn
requires a more cautious contraction criterion leading to larger numbers of coefficients
in the series representation for $p(\phi)$, and therefore longer computation times are
expected. If the goal of using subsets is to lower computation times for determining the
optimal $k$, the potential increase in computation due to requiring more coefficients must
therefore also be considered.

The results for the multi-step method show that while it performs much better when the
compared expected gains are for sequences that require exactly the same time (rather than
approximately), it does not perform particularly better than the simpler and more
versatile hybrid gain rate method. In particular, only the multi-step method optimising
entropy gain reaches HS, but with a notably larger pre-factor of 2.13 compared to the gain
rate methods.

For the estimation sequences using a total time of $2^{16}$ a summary of some computation
time benchmarks for gain rate methods are given in table \ref{tab:benchmarks}. Since for
later shots in the sequence contraction is also performed on a significant fraction of
shots, times required for this computation are also listed. Here we have studied the cases
where $k \in \{2^n\}\, , n = 0,1,2, \dots$, while the cases where all $k$-values (up to
some maximum value) are searched using a series of Fibonacci searches was studied in the
main text. For comparison the case where a brute force search is performed over all
$k$-values is also included.

We observe that the time required for the Bayesian update depends on the method used. We
expect this to be related to the size of $k$-values chosen by a strategy since this
determines how many coefficients must be used in the Fourier representation. Optimising
the rate of entropy gain tends to choose larger $k$-values than the sharpness. When
considering only a subset of possible $k$-values, we expect that occasionally the
algorithm will choose a larger value of $k$ than if all values were considered. This would
then have the effect of increasing \emph{maximum} update time. We see that this is indeed
the case when restricting to the subset $k \in \{2^n\}\, , n = 0,1,2, \dots$, and that the
\emph{average} update time is also slightly longer.

When considering the times required to determine optimal values of $\alpha$ and $k$, it's
important to note that since the optimal $k$ increases with phase knowledge, later shots
(i.e. measurements) generally use higher $k$-values and require more time for the
optimisation. When the phase knowledge is sufficient, contractions are performed that
prevent the required computation per shot from increasing further. In the hybrid method,
at most the first half of the total time (for values in table \ref{tab:benchmarks},
$2^{15}$) is used for shots with $\alpha$ and $k$ optimised using entropy gain rate; since
$k$ tends to increase, this is most of the shots in the estimation sequence. For the
remaining time $\alpha$ and $k$ are chosen to maximise the expected rate of sharpness
gain. Due to these changes in the computation time with every shot, the times listed in
table \ref{tab:benchmarks} for the hybrid method cannot be directly compared with those
where the expected rate of sharpness gain is maximised for all shots.

Using the same implementation and similar processor in an experiment we see that the brute
force optimisation over all $k$ values would be appropriate for single shot times of at
least $\sim 10\; m s$. Using a series of Fibonacci searches, as was done for the methods
studied in the main text, would be appropriate for single shot times of at least
$\sim 1\; m s$ for the hybrid method, and $\sim 500\; \mu s$ when using only sharpness
gain. Using the subset $k \in \{2^n\}\, , n = 0,1,2, \dots$ would be appropriate for
experiments with times of at least $\sim 400\; \mu s$ for the hybrid method, and
$\sim 100\; \mu s$ when using only sharpness gain.

\section{\label{sec:sup-int-time-rep} Intermediate time-repetition regime}

In the main text we discussed the performance of TAPE in different settings where the
relation between the time of an experiment and the number of repetitions $k$ of the
unknown phase (coherent evolution) is different. In particular we considered two extreme
cases: $\mathbf{T} \propto \mathbf{K}$ which is the usual situation in quantum metrology,
and $t = 1\, \forall\, k$ which is practically relevant for experiments where the
measurement time is several orders of magnitude greater than the time of coherent
evolution. Here we consider an intermediate case where the time required for state
preparation and measurement (SPAM) is 100 times greater than the time required for a
single application of the unknown phase ($k = 1$). In addition we assume a system with
decoherence as described in the main text, i.e. perfectly prepared $\ket{+}$ states,
$U = \Exp{-iZ\phi/2}$, $Z$, and each application $U$ of the unknown phase is followed by a
dephasing channel $K_0 = \sqrt{\frac{1+\eta}{2}}\openone$,
$K_1 = \sqrt{\frac{1-\eta}{2}}Z$ with $\eta = 0.995$.
\begin{figure}[h]
  \centering
  \includegraphics[width=0.48\textwidth]{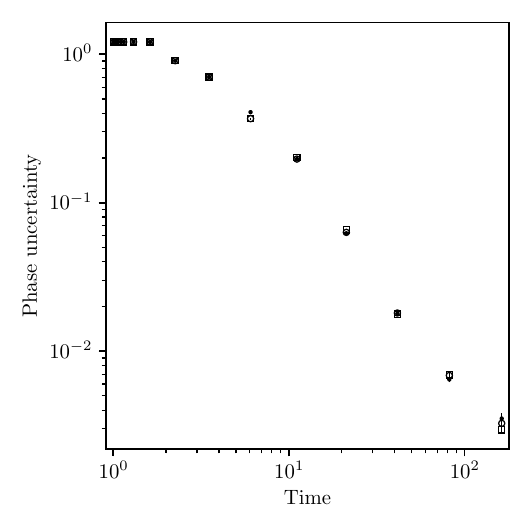}
  \caption{Phase uncertainty of estimates ($10^6$ realisations of estimation) obtained by
    applying TAPE in the case that the time required for SPAM is 100 times greater than
    the time required for a single application of the unknown phase, and in the presence
    of dephasing ($\eta = 0.995$). In particular, the time on the $x$-axis of the plot is
    related to the number of applications $k$ of the unknown phase by $t = (k +
    100)/101$. The correct dephasing rate is included in the model of the
    estimators. Expected entropy gain: $\bullet$, expected sharpness gain: $\square$,
    hybrid method: $\circ$.}
  \label{fig:intermediate-scaling}
\end{figure}

The results of $10^6$ realisations of estimation using the gain rate method with entropy
gain, sharpening gain, and the hybrid method are plotted in figure
\ref{fig:intermediate-scaling}. For times shorter than twice the SPAM time the phase
uncertainty decreases little because we can only perform one measurement. For times
slightly more than twice the SPAM time the situation is similar to the case
$t = 1\, \forall\, k$ discussed in the main text since applying large values of $k$ does
not significantly impact the experiment time unless $k \gg 100$. In the large $k$ limit we
approach the usual situation in quantum metrology $\mathbf{T} \propto \mathbf{K}$, however
since in this case $k$ is limited by decoherence we never reach this limit. All three of
the plotted methods preform similarly. Initially this is consistent with the results of
figure \ref{fig:shot-scaling} for short times. In the large $k$ limit the situation
becomes similar to the situation in figure \ref{fig:decoherence-scaling} for longer times,
where all three methods perform similarly when decoherence is included in the model of the
estimator. Thus, we see that the results in the limiting cases of the relation between
$\mathbf{T}$ and $\mathbf{K}$ provide an indication of each method's performance in the
intermediate case.

\section{\label{sec:sup-multi-step} Multi-step gain method}

Supposing we have a prior probability density $p_{s-1}(\phi)$ at step $s-1$ of a sequence
of measurements, we calculate the maximum expected knowledge (sharpness or entropy) gain
for step $s$ as $\Delta G(s) \equiv \max_{\alpha_s, k_s} \Delta G(\alpha_s, k_s)$ for a
single measurement using \eqref{eq:sharpness-gain} or \eqref{eq:entropy-gain} for the
sharpness or entropy gain, respectively. We calculate the maximum expected gain for two
measurements as
\begin{align*}
  \Delta G(s,s+1) &\equiv \Delta G(s) + \sum_{\xi}
                    \Pi_{\xi}(\alpha_{s+1}^{(\xi)},k_{s+1}^{(\xi)}) \Delta G(s+1) \, .
\end{align*}
Since $\Delta G(\alpha_{s+1},k_{s+1})$ depends on $p_s(\phi|\xi; \alpha_s,k_s)$, it
implicitly depends on the outcome $\xi$. Therefore we write $\alpha_{s+1}^{(\xi)}$,
$k_{s+1}^{(\xi)}$ to denote the values optimised conditioned on a particular measurement
outcome at step $s$. We can similarly calculate the maximum expected gain for three
measurements as
\begin{align*}
  &\Delta G(s,s+2) \equiv \\
  &\Delta G(s) + \sum_{\xi}
    \Pi_{\xi}(\alpha_{s+1}^{(\xi)},k_{s+1}^{(\xi)}) \Delta G_{\xi}(s+1,s+2) \, ,
\end{align*}
where the subscript $\xi$ in $\Delta G_{\xi}(s+1,s+2)$ is to remind us that this
quantity depends on the measurement outcome at step $s$. Applying this recursively we can
write the maximum expected gain for $j$ measurements as
\begin{align}
  &\Delta G(s,s-1+j) = \Delta G(s) \nonumber \\
  &+ \sum_{\xi}
    \Pi_{\xi}(\alpha_{s+1}^{(\xi)},k_{s+1}^{(\xi)}) \Delta G_{\xi}(s+1,s-1+j) \, .
    \label{eq:multi-step-gain}
\end{align}
Since the number of possible outcomes for the next $j$ measurements is $2^j$, the
computation of the multi-step gain generally grows exponentially with the number of
measurements. For this reason algorithms \ref{alg:intervals} and \ref{alg:compare} are
written to compare measurement sequences of approximately equal time that contain at most
5 measurements. In general the gain calculations \eqref{eq:sharpness-gain},
\eqref{eq:entropy-gain} don't require all coefficients in \eqref{eq:density} allowing a
more efficient computation of the multi-step gain.

\begin{algorithm}[H]
  \caption{Calculate intervals}
  \label{alg:intervals}
  \begin{algorithmic}
    \State Input: $\mathbf{T} = \{t_1, t_2, \dots , t_N\},\, t_n \in \mathbb{R}$, sorted
    in increasing order.

    \Function{Intervals}{$\mathbf{T}$}
    \State Let $\mathbf{I}$ be a list of pairs $(a,b)$. $a,b \in \mathbb{N}$.
    \State\Comment $\mathbf{I}.\text{push}((a,b))$ denotes adding a pair $(a,b)$ to
    the back of the list.
    \State $\mathbf{I}$ is initially empty.
    \State $n \gets 1$
    \While {$n < N$}
    \State $m \gets n$
    \State $n \gets n + 1$
    \While {$n \le N$ and $t_n/t_m < 8/7$}
    \State $n \gets n + 1$
    \EndWhile
    \State $\mathbf{I}.\text{push}((m,n-1))$
    \EndWhile
    \If {$n - 1 < N$}
    \State $\mathbf{I}.\text{push}((n,N))$
    \EndIf
    \State \Return $\mathbf{I}$
    \EndFunction
  \end{algorithmic}
\end{algorithm}
\begin{algorithm}[H]
  \caption{Compare two $k$-values}
  \label{alg:compare}
  \begin{algorithmic}
    \State Requires: function gain$(k,j)$, where $k$ is the number of applications of the
    unitary $U$ and $j$ is the number of steps. gain$(k,j)$ returns the expected knowledge
    gain for performing $j$ steps (measurements) with $U^k$, i.e. the result of equation
    \eqref{eq:multi-step-gain} with $k_s = k_{s+1} = \dots = k_{s-1+j} = k$.

    \State Inputs:
    \State $\mathbf{K} = \{k_1, k_2, \dots , k_N\},\, k_n \in \mathbb{N}$,
    $\mathbf{T} = \{t_1, t_2, \dots , t_N\},\, t_n \in \mathbb{R}$,
    \State $n_1, n_2 \in \mathbb{N}$, $n_1 < n_2$, $t_{n_1}/t_{n_2} < 32/7$.
    \Function{Compare}{$\mathbf{K}$, $\mathbf{T}$, $n_1$, $n_2$}
    \State $r \gets t_{n_2}/t_{n_1}$
    \If {$r < 8 / 7$}
    $j_1 \gets 1$, $j_2 \gets 1$
    \ElsIf {$r < 24 / 17$}
    $j_1 \gets 4$, $j_2 \gets 3$
    \ElsIf {$r < 12 / 7$}
    $j_1 \gets 3$, $j_2 \gets 2$
    \ElsIf {$r < 20 / 9$}
    $j_1 \gets 2$, $j_2 \gets 1$
    \ElsIf {$r < 30 / 11$}
    $j_1 \gets 5$, $j_2 \gets 2$
    \ElsIf {$r < 24 / 7$}
    $j_1 \gets 3$, $j_2 \gets 1$
    \ElsIf {$r < 32 / 7$}
    $j_1 \gets 4$, $j_2 \gets 1$
    \EndIf
    \State $g_1 \gets \text{gain}(k_{n_1}, j_1)$, $g_2 \gets \text{gain}(k_{n_2}, j_2)$
    \If {$g_1 > g_2$} \Return $n_1$
    \Else {} \Return $n_2$
    \EndIf
    \EndFunction
  \end{algorithmic}
\end{algorithm}
\begin{algorithm}[H]
  \caption{Search interval}
  \label{alg:search-interval}
  \begin{algorithmic}
    \State Inputs:
    \State $\mathbf{K} = \{k_1, k_2, \dots , k_N\},\, k_n \in \mathbb{N}$,
    $\mathbf{T} = \{t_1, t_2, \dots , t_N\},\, t_n \in \mathbb{R}$,
    \State $n_1, n_2 \in \mathbb{N}$, $n_1 \le n_2$
    \Function {searchInterval}{$\mathbf{K}$, $\mathbf{T}$, $n_1$, $n_2$}
    \If {$n_2 == n_1$}
    \State \textbf{if} $n_1 == N$ \textbf{then} \Return $n_1$
    \State \Return compare$(n_1, n_1 + 1)$
    \Else \Comment Brute-force search for max gain in the interval.
    \State $\text{max} \gets \text{gain}(k_{n_1}, 1)$,
    $n \gets n_1$
    \For {$m \gets n_1 + 1, n_2$}
    \State $g \gets \text{gain}(k_m, 1)$
    \If {$g > \text{max}$}
    \State $\text{max} \gets g$, $n \gets m$
    \EndIf
    \EndFor
    \State \Return $n$
    \EndIf
    \EndFunction
  \end{algorithmic}
\end{algorithm}
\begin{algorithm}[H]
  \caption{Search up}
  \label{alg:search-up}
  \begin{algorithmic}
    \State Requires: function findInterval$(\mathbf{I},n)$, where $\mathbf{I}$ is a list
    of intervals (see algorithm \ref{alg:intervals}) and $n \in \mathbb{N}$, that returns
    the index $i$ of the \emph{unique} interval in $\mathbf{I}$ s.t. $n1 \le n \le n2$.

    \State Inputs:
    \State $\mathbf{K} = \{k_1, k_2, \dots , k_N\},\, k_n \in \mathbb{N}$,
    $\mathbf{T} = \{t_1, t_2, \dots , t_N\},\, t_n \in \mathbb{R}$,
    \State intervals $\mathbf{I}$ (see algorithm \ref{alg:intervals}),
    $N_I$ (the number of intervals in $\mathbf{I}$),
    $n \in \mathbb{N}$
    \Function{searchUp}{$\mathbf{K}$, $\mathbf{T}$, $\mathbf{I}$, $n$}
    \State $i \gets \text{findInterval}(\mathbf{I},n)$
    \While {\texttt{true}}
    \State $(n_1, n_2) \gets \mathbf{I}[i]$
    \State $n \gets \text{searchInterval}(\mathbf{K}, \mathbf{T}, n_1, n_2)$
    \If {$n_2 == n_1$}
    \State \textbf{if} $n == n_1$ \textbf{then} \Return $n$
    \Else
    \State \textbf{if} $n < n_2$ \textbf{then} \Return $n$
    \State \textbf{if} $i == N_I$ \textbf{then} \Return $n$
    \State $m \gets \text{compare}(\mathbf{K}, \mathbf{T}, n_1, n_2 + 1)$
    \State \textbf{if} $m == n_1$ \textbf{then} \Return $n_1$
    \EndIf
    \State $i \gets i + 1$
    \EndWhile
    \EndFunction
  \end{algorithmic}
\end{algorithm}
\begin{algorithm}[H]
  \caption{Multi-step gain method}
  \label{alg:multi-step-method}
  \begin{algorithmic}
    \State Inputs:
    \State $\mathbf{K} = \{k_1, k_2, \dots , k_N\},\, k_n \in \mathbb{N}$,
    $\mathbf{T} = \{t_1, t_2, \dots , t_N\},\, t_n \in \mathbb{R}$,
    \Function {multiStepMethod}{$\mathbf{K}, \mathbf{T}$}
    \State $\mathbf{I} \gets \text{intervals}(\mathbf{T})$
    \State $n \gets \text{searchUp}(\mathbf{K}, \mathbf{T}, \mathbf{I}, 1)$
    \State \Return $k_n$
    \EndFunction
  \end{algorithmic}
\end{algorithm}

\begin{figure}[h]
  \centering
  \includegraphics[width=0.5\textwidth]{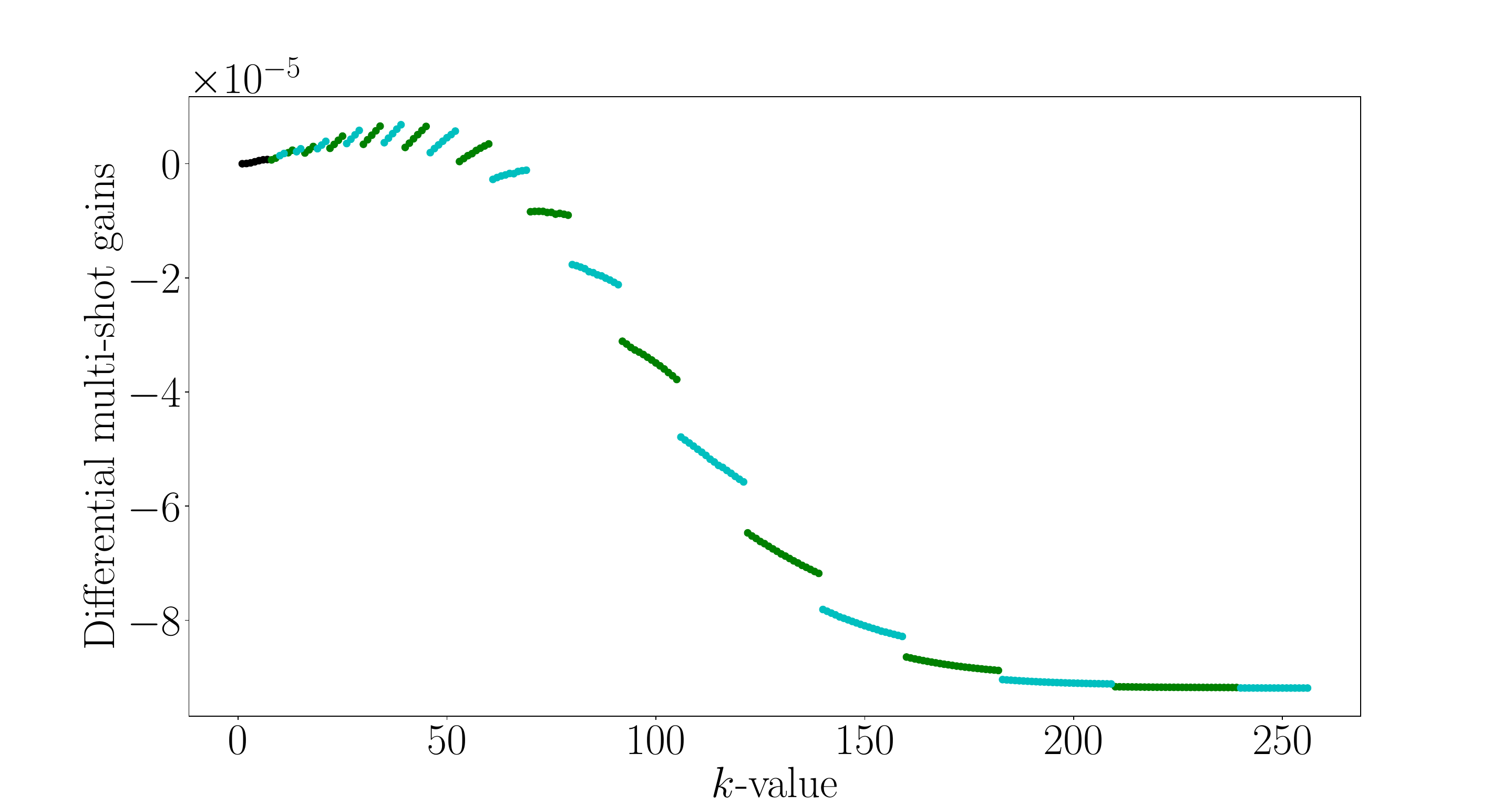}
  \includegraphics[width=0.5\textwidth]{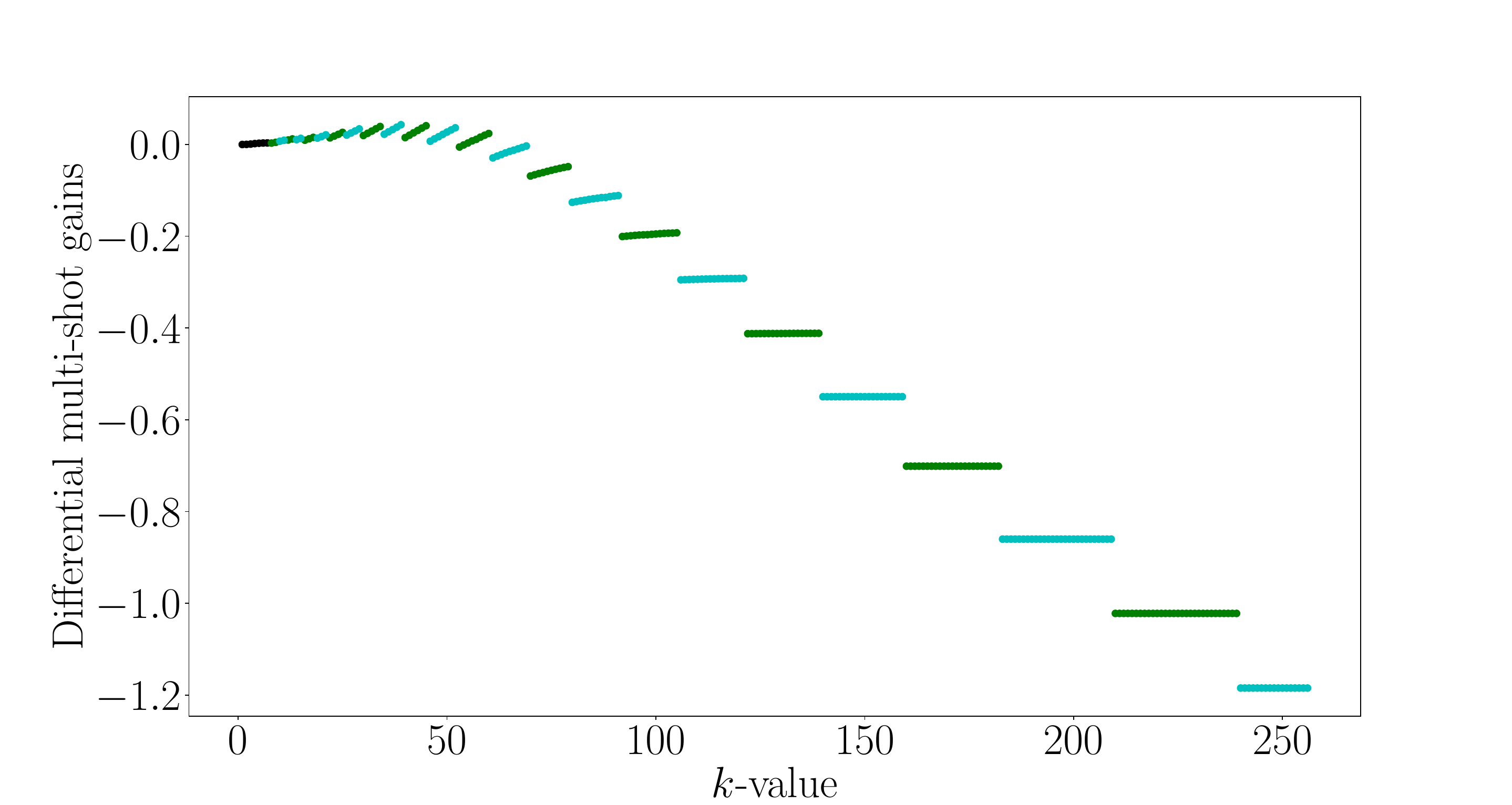}
  \caption{Differential gains for the multi-step method. Upper (Lower) plot: differential
    multi-shot gains for the expected sharpness (entropy) gain. The expected sharpness
    becomes flat for larger values of $k$ since measurements with such large $k$-values
    would lead to multi-peaked densities for the phase knowledge giving no increase in
    sharpness. Since multi-peaked densities still lead to a change in entropy, the
    differential gains in the lower plot continue to decrease for larger $k$-values. The
    intervals determined by algorithm \ref{alg:intervals} are shown by the colours. Black
    points correspond to intervals containing only one $k$-value. Intervals with more than
    one $k$-value are coloured, with a change in colour indicating the next interval.}
  \label{fig:diff-gains}
\end{figure}

To understand the effective functions maximised by the multi-step method (in the
noise-free case), we calculate the multi-shot gains for sequences with approximately equal
time as detailed in algorithm \ref{alg:compare}. We assume the usual experiment times in
sequential metrology experiments, i.e. $\mathbf{T} \propto \mathbf{K}$, and perform
estimation using a total time of $\sum_s k_s \approx 768$ starting from a uniform prior.
In figure \ref{fig:diff-gains} we then plot the cumulative difference in expected gain for
the next measurement between consecutive $k$-values:
\begin{align}
  \sum_{m=2}^{k} \big(\mathrm{gain}(m,j_m) - \mathrm{gain}(m-1,j_{m-1})\big)\, ,
  \label{eq:diff-gain-sum}
\end{align}
where the values of $j_m,j_{m-1} \forall m = 2,\dots , k$ are determined as in algorithm
\ref{alg:compare}. The gain difference calculated in \eqref{eq:diff-gain-sum} is adjusted
between intervals with multiple $k$-values by comparing the first $k$-value of consecutive
intervals (as in algorithm \ref{alg:search-up}). The values plotted for $k=1$ are set to
zero. In this way the plotted values show an effective function that is maximised
(locally) by algorithm \ref{alg:multi-step-method}.

We see that the differential gains for both expected sharpness and entropy gains show a
global maximum around $k \sim 45$. However the differences in multi-shot gains oscillate
due to the fact that the times of the multi-step sequences are only approximately
equal. Due to these oscillations, algorithm \ref{alg:search-up} tends to stop at a local
maximum. In algorithm \ref{alg:multi-step-method} we always start the search at $k=1$ so
that the value of $k$ returned is typically much lower than the optimal value suggested by
the plotted differential gains. Using the gain rate method to determine the choice of $k$
for this measurement gives $50 (60)$ to maximise the expected sharpness (entropy) gain
rate. More efficient methods to find the global maximum of the plotted functions could
potentially also lead to accurate estimation procedures.

\section{\label{sec:sup-cnt-cpx} Computational complexity of contractions}

In this section we derive an upper bound on the number of coefficients $\Gamma$
needed for the Fourier series $p(\phi)$ when using the contraction method, under the
assumption that $p(\phi)$ is Gaussian. In particular, since $p(\phi)$ is periodic, we
assume that it would have the form
\begin{align*}
  p(\phi) &= \sum_{n=-\infty}^{\infty} \mathrm{e}^{-\frac{1}{2}n^2 \sigma^2}
            \mathrm{e}^{i n (\phi - \mu)} \\
          &= 1 + 2\sum_{n=1}^{\infty} \mathrm{e}^{-\frac{1}{2}n^2 \sigma^2}\cos(n(\phi-\mu)) \, ,
\end{align*}
where $\mu$ is the mean phase, and the Holevo variance is $V[p(\phi)] = \sigma^2$. We have
written the sum to $\infty$ here; below we will consider the number of coefficients
$\Gamma$ that would be used in an estimation sequence.

In the following we analyse the case where we perform contractions with $m=2$, so that
after performing $c$ contractions we go from a representation $p(\phi)$ on an interval of
size $2\pi$ to a representation $q(\theta)$ on an interval of size $2\pi / 2^c$. Since
$p(\phi)$ is a measure for the probability of the system phase we are trying to estimate,
we will quantify the probability of the system phase being outside the interval
represented by $q(\theta)$ after a single contraction by
\begin{align*}
  \varepsilon &= \int_{\mu + \frac{\pi}{2}}^{\mu + \frac{3\pi}{2}} \frac{p(\phi)}{2\pi} d\phi \\
              &= \frac{1}{2\pi} \int_{\frac{\pi}{2}}^{\frac{3\pi}{2}}
                \left(1 + 2\sum_{n=1}^{\infty} \mathrm{e}^{-\frac{1}{2}n^2 \sigma^2}
                \cos(n\phi)\right)d\phi \\
              &= \frac{1}{2} + \frac{1}{\pi} \sum_{n=1}^{\infty} \mathrm{e}^{-\frac{1}{2}n^2 \sigma^2}
                \int_{\frac{2}{\pi}}^{\frac{3\pi}{2}} \cos(n\phi)d\phi \\
              &= \frac{1}{2} + \frac{2}{\pi}\sum_{n=0}^{\infty}
                \frac{\mathrm{e}^{-\frac{1}{2}\sigma^2 (4n+3)^2}}{4n+3} -
                \frac{\mathrm{e}^{-\frac{1}{2}\sigma^2 (4n+1)^2}}{4n+1} \, ,
\end{align*}
where in the second line we have used the fact that the result is independent of the mean
phase $\mu$ so that we can, without loss of generality, set $\mu$ to zero. The last line
follows from solving the integral and relabelling the index of the sum to include only the
odd values of $n$ in the third line (the integral is zero for even values of
$n$). Although we have not found a solution to the infinite series, we find that
$\varepsilon$ is very close to the corresponding quantity for a non-periodic Gaussian,
\begin{align*}
  \varepsilon \approx 1 + \mathrm{erf}\left(\frac{-\pi}{2\sqrt{2}\sigma}\right) \, ,
\end{align*}
when $\sigma$ is less than $\sim \pi/2$.

If we specify an upper bound $\varepsilon$ on the desired probability that the system
phase cannot be represented after performing a contraction, then we obtain an upper bound
on the standard deviation, $\sqrt{V[p(\phi)]} = \sigma$, before we should perform a
contraction:
\begin{align}
  \sigma < \frac{\pi}{2\sqrt{2}\mathrm{erf}^{-1}(1-\varepsilon)}
  = \frac{\pi}{2\sqrt{2}\mathrm{erfc}^{-1}(\varepsilon)} \, ,
  \label{eq:std-err}
\end{align}
where $\mathrm{erfc}^{-1}$ is the inverse complementary error function. As mentioned in
the main text, we expect from Fourier analysis that the number of coefficients
$\Gamma$ is inversely proportional to $\sqrt{V[p(\phi)]}$ (and therefore to
$\Delta\hat{\phi}$). We can see this in the case of a Gaussian by assuming that any
numerical representation will have a finite machine precision. Let $\delta$ be the
smallest value that can be represented numerically. Then for the highest order coefficient
that can be represented, we have $\mathrm{e}^{-\Gamma^2\sigma^2 / 2} = \delta$, so
\begin{align}
  \Gamma &= \frac{\sqrt{2 \ln (\delta^{-1})}}{\sigma} \, .
                \label{eq:N-std}
\end{align}
Substituting \eqref{eq:std-err} into \eqref{eq:N-std}, we obtain a relation between the
probability of an error, $\varepsilon$, and the number of coefficients:
\begin{align}
  \Gamma &> \frac{4}{\pi}\sqrt{\ln(\delta^{-1})}\mathrm{erfc}^{-1}(\varepsilon) \, .
                \label{eq:N-err}
\end{align}

If we now have a $p(\phi)$ that is Gaussian with standard deviation $\sigma^*$ given by
\eqref{eq:std-err} and that requires a representation to machine precision using
$\Gamma^*$ coefficients, as given by \eqref{eq:N-err}, then the probability of an
error after performing one contraction with $m=2$ will be less than $\varepsilon$. At this
point we would have a density $q(\theta)$ requiring only $\Gamma^*/2$ coefficients
and we could continue estimation until the standard deviation of $q(\theta)$ would be
$\sigma^*$, and that of $p(\phi)$ represented by this contraction would be $\sigma^*/2$.
Now $q(\theta)$ would have $\Gamma^*$ coefficients and we would need to perform
another contraction to avoid increasing $\Gamma$ above $\Gamma^*$. If we would
perform a second contraction the total probability that \emph{either} the first or the
second contraction would lead to an error would be
$\varepsilon + \varepsilon(1 - \varepsilon) < 2\varepsilon$. To keep the total error
probability below the original value of $\varepsilon$ we would then want to use slightly
more coefficients to begin with so that $\varepsilon$ is twice as small.

\begin{figure}[h]
  \centering
  \includegraphics[width=0.5\textwidth]{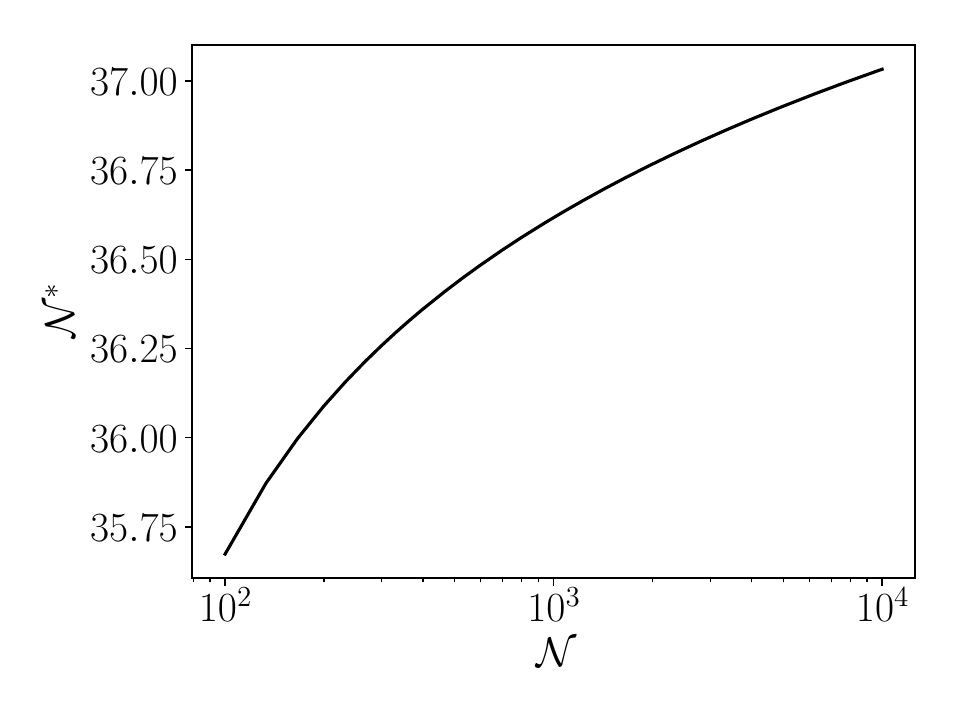}
  \caption{Plot of the upper bound \eqref{eq:Nstar-of-N} with $\epsilon = 10^{-10}$ and
    $\delta = 10^{-16}$.}
  \label{fig:contraction-bound-HS}
\end{figure}

More generally, if we perform a total of $c$ contractions, the total error probability
$\epsilon$ is less than $c\varepsilon$, and from \eqref{eq:N-err} we should use a number
of coefficients
\begin{align*}
  \Gamma^* &\equiv \left\lceil \frac{4}{\pi}\sqrt{\ln(\delta^{-1})}
                  \mathrm{erfc}^{-1} \left(\frac{\epsilon}{c}\right) \right\rceil \, .
\end{align*}
A representation without contractions would require $\Gamma = 2^c \Gamma^*$
coefficients. Let $\Gamma_1$ be the number of coefficients required to keep the error
probability below $\epsilon$ for a single contraction:
\begin{align*}
  \Gamma_1 &\equiv \left\lceil \frac{4}{\pi}\sqrt{\ln(\delta^{-1})}
                  \mathrm{erfc}^{-1} \left(\epsilon\right) \right\rceil \, .
\end{align*}
We have $\Gamma = 2^c \Gamma^* \geq 2^c \Gamma_1$, so
$c \leq \log_2(\Gamma/\Gamma_1)$, and we can write an upper bound on the number
of coefficients $\Gamma^*$ when using contractions as a function of the number of
coefficients $\Gamma$ that would be needed without using contractions:
\begin{align}
  \Gamma^* \leq \frac{4}{\pi}\sqrt{\ln(\delta^{-1})}
  \mathrm{erfc}^{-1}\left(
  \frac{\epsilon}{\log_2\left(\frac{\Gamma}{\Gamma_1}\right)}
  \right) \, .
  \label{eq:Nstar-of-N}
\end{align}
In figure \ref{fig:contraction-bound-HS} we plot the upper bound \eqref{eq:Nstar-of-N}
with $\epsilon = 10^{-10}$ and $\delta = 10^{-16}$. We see that it scales slower than
$\log(\Gamma)$ suggesting that the we can use a number of coefficients
$\Gamma^* = \mathcal{O}(\log(\Gamma))$ while keeping the total probability that
an error occurs below $\epsilon$.

To determine the asymptotic complexity of $\Gamma^*$ in terms of $\Gamma$ we
make use of an asymptotic expansion of $\mathrm{erfc}^{-1}(x)$, for $x \rightarrow 0$
\cite{2024dlmfnist, 1976blair}:
\begin{align}
  \mathrm{erfc}^{-1}(x) \sim u^{-1/2} \left( 1 +
  u \left(a_2 u + a_3 u^2 + \dots \right) \right)
  \label{eq:inverfc-asym}
\end{align}
where
\begin{align*}
  u &= -2/\ln(\pi x^2 \ln(1/x)) \, , \\
  a_2 &= v/8 \, , \qquad a_3 = -(v^2 + 6v - 6)/32 \, , \\
  v &= \ln(\ln(1/x)) - 2 + \ln(\pi) \, .
\end{align*}
As $x \rightarrow 0$, $v \rightarrow \infty$ and $u \rightarrow 0$, but each term
$a_{n+1} u^n$ in \eqref{eq:inverfc-asym} will be $\mathcal{O}\left((vu)^n\right)$. Since
\begin{align*}
  \lim_{x \rightarrow 0} vu \rightarrow 0 \, ,
\end{align*}
$\mathrm{erfc}^{-1}(x) \rightarrow u^{-1/2}$ as $x \rightarrow 0$. Then to go further, as
$x \rightarrow 0$, we have $u^{-1/2} < \sqrt{\ln(1/x)}$. From this result we have
$\Gamma^* = \mathcal{O}\left(\sqrt{\log(\log(\Gamma))}\right)$.

We note that the bound \eqref{eq:Nstar-of-N} is only of practical interest for determining
what value of $\sigma$ should be reached before performing a contraction with a certain
error probability, if $p(\phi)$ is Gaussian. Since we have not shown this in any of the
estimation methods we have studied in the main text it cannot be used there. In those
cases the value of $\sigma$ to be reached before contracting is determined empirically
from simulation. But the purpose of the bound \eqref{eq:Nstar-of-N} is mainly to determine
what potential the method of using contractions has to reduce computational complexity,
rather than to be a practical way of calculating the required value of $\sigma$.

\onecolumngrid


\section{\label{sec:sup-ent-gain} Expected entropy gain}

In this section we show the relation between the expected entropy gain
\eqref{eq:entropy-gain-def} and the Kullback-Leibler divergence (KL divergence), and we
derive equation \eqref{eq:entropy-gain}. Starting from the definition of the expected
entropy gain \eqref{eq:entropy-gain-def}, we can rewrite the entropy of the posterior (we
sometimes write $p_s(\phi) = p_s(\phi | \xi ; \alpha, k)$ for short)
\begin{align*}
  H[p_s(\phi| \xi; \alpha, k)]
  &= -\int_0^{2\pi} \frac{d\phi}{2\pi} p_s(\phi) \ln\left(\frac{P_{\xi}(\alpha, k\phi)}{\Pi_{\xi}(\alpha,k)}\right)
    -\int_0^{2\pi} \frac{d\phi}{2\pi} p_s(\phi) \ln\left(\frac{p_{s-1}(\phi)}{2\pi}\right) \\
  &= -\int_0^{2\pi} \frac{d\phi}{2\pi} p_s(\phi) \ln\left(\frac{p_s(\phi)}{p_{s-1}(\phi)}\right)
    + H(p_s ,\, p_{s-1}) \\
  &= -D_{KL}(p_s \| p_{s-1}) + H(p_s ,\, p_{s-1}) \, ,
\end{align*}
where $D_{KL}(p_s\| p_{s-1})$ is the \emph{KL divergence} of the posterior from the prior,
and $H(p_s ,\, p_{s-1})$ is the \emph{cross entropy}. Noting that
\begin{align*}
  \sum_{\xi} \Pi_{\xi}(\alpha, k)
  \Big( H[p_{s-1}(\phi)] - H\big(p_s(\phi|\xi;\alpha,k) ,\, p_{s-1}(\phi)\big) \Big)
  &= 0 \, ,
\end{align*}
we see that
\begin{align*}
  \Delta_s H(\alpha,\, k)
  &= \sum_{\xi} \Pi_{\xi}(\alpha, k) D_{KL}(p_s\| p_{s-1}) \, ,
\end{align*}
i.e. the expected entropy gain is equal to the expected KL divergence. We can rewrite this
as
\begin{align*}
  \Delta_s H(\alpha,\, k) &= \sum_{\xi} \int_0^{2\pi} \frac{d\phi}{2\pi} P_{\xi}(\alpha , k\phi)p_{s-1}(\phi)
                            \ln \big(P_{\xi}(\alpha , k\phi)\big) -
                            \sum_{\xi} \Pi_{\xi}(\alpha,\, k) \ln \left(\Pi_{\xi}(\alpha,\, k)\right) \, .
\end{align*}
We would like to rewrite the first term with the integral in a form that can be computed
more efficiently.

\subsection{Solving the integral}
We have
\begin{align*}
  &\sum_{\xi} \int_0^{2\pi} \frac{d\phi}{2\pi} P_{\xi}(\alpha , k\phi)p_{s-1}(\phi)
    \ln \big(P_{\xi}(\alpha , k\phi)\big) \\
  &= -\ln(2) + \sum_{\xi} \int_0^{2\pi} \frac{d\phi}{2\pi} \bigg[
    \frac{1}{2} \Big(
    1 + \xi \big( (1-\lambda) + \lambda\zeta\cos(\alpha - k\phi) \big)
    \Big) \\
  &\qquad\qquad\qquad\qquad\qquad\quad \ln\Big(
    1 + \xi \big( (1-\lambda) + \lambda\zeta\cos(\alpha - k\phi) \big)
    \Big)
    \bigg] \, .
\end{align*}
Let $\gamma = \zeta\cos(\alpha - k\phi)$, $\beta = (1-\lambda) + \lambda\gamma$. Then we
have
\begin{align*}
  &\sum_{\xi} \int_0^{2\pi} \frac{d\phi}{2\pi} P_{\xi}(\alpha , k\phi)p_{s-1}(\phi)
    \ln \big(P_{\xi}(\alpha , k\phi)\big) \\
  &= -\ln(2) + \int_0^{2\pi} \frac{d\phi}{2\pi} \frac{1}{2} \bigg[
    \Big(\ln(1+\beta) + \ln(1-\beta)\Big)
    + \beta\Big(\ln(1+\beta) - \ln(1-\beta)\Big)
    \bigg] \, .
\end{align*}
We can work out the terms in the last line using the Taylor series form of
$\ln(1 \pm \beta)$ \footnote{This is possible since $|\beta| \le 1$.}:
\begin{align*}
  \ln(1+\beta) + \ln(1-\beta)
    &= -2 \sum_{n = 1}^\infty \frac{\beta^{2n}}{2n} \, ,
\end{align*}
\begin{align*}
  \ln(1+\beta) - \ln(1-\beta)
  &= 2 \sum_{n = 1}^\infty \frac{\beta^{2n-1}}{2n-1} \, .
\end{align*}
So we have
\begin{align*}
  \Big(\ln(1+\beta) + \ln(1-\beta)\Big)
  + \beta\Big(\ln(1+\beta) - \ln(1-\beta)\Big)
  &= 2 \sum_{n=1}^\infty \left(\frac{1}{2n(2n-1)}\right)\beta^{2n} \, .
\end{align*}
Using this result, we have
\begin{align*}
  \sum_{\xi} \int_0^{2\pi} \frac{d\phi}{2\pi} P_{\xi}(\alpha , k\phi)p_{s-1}(\phi)
  \ln \big(P_{\xi}(\alpha , k\phi)\big) &= -\ln(2) + \sum_{n=1}^\infty \left( \frac{1}{2n(2n-1)} \right)
                                              \int_0^{2\pi} \frac{d\phi}{2\pi} p_{s-1}(\phi) \beta^{2n} \, ,
\end{align*}
and
\begin{align*}
  \beta^{2n} &= \big( (1-\lambda) + \lambda\gamma \big)^{2n}
               = \sum_{u=0}^{2n}
               \begin{pmatrix}
                 2n \\
                 u
               \end{pmatrix}
  (1-\lambda)^{2n-u}\lambda^u \gamma^u \, ,
\end{align*}
so
\begin{align*}
  &\sum_{\xi} \int_0^{2\pi} \frac{d\phi}{2\pi} P_{\xi}(\alpha , k\phi)p_{s-1}(\phi)
    \ln \big(P_{\xi}(\alpha , k\phi)\big) \\
  &= -\ln(2) + \sum_{n=1}^\infty \left( \frac{1}{2n(2n-1)} \right)
    \sum_{u=0}^{2n}
    \begin{pmatrix}
      2n \\
      u
    \end{pmatrix}
  (1-\lambda)^{2n-u}\lambda^u \zeta^u
  \int_0^{2\pi} \frac{d\phi}{2\pi} p_{s-1}(\phi) \cos^u(\alpha - k\phi) \, .
\end{align*}
We can rewrite the integral term
\begin{align*}
  \int_0^{2\pi} \frac{d\phi}{2\pi} p_{s-1}(\phi) \cos^u(\alpha - k\phi)
  &= \frac{1}{2^u} \sum_{q=0}^u
    \begin{pmatrix}
      u \\
      q
    \end{pmatrix}
  \Exp{i\alpha(u-2q)} c_{k(u-2q)}^{(s-1)} \, ,
\end{align*}
so
\begin{align}
  &\sum_{\xi} \int_0^{2\pi} \frac{d\phi}{2\pi} P_{\xi}(\alpha , k\phi)p_{s-1}(\phi)
    \ln \big(P_{\xi}(\alpha , k\phi)\big) = \nonumber \\
  &-\ln(2) + \sum_{n=1}^\infty \left( \frac{1}{2n(2n-1)} \right) \sum_{u=0}^{2n}
    \begin{pmatrix}
      2n \\
      u
    \end{pmatrix}
  (1-\lambda)^{2n-u} \left(\frac{\lambda \zeta}{2}\right)^u \sum_{q=0}^u
  \begin{pmatrix}
    u \\
    q
  \end{pmatrix}
  \Exp{i\alpha(u-2q)} c_{k(u-2q)}^{(s-1)} \, .
  \label{eq:int-sol}
\end{align}
That solves the integral, but we are left with some infinite series that are not very
practical for computation. Ideally we would like to have an expression that is a sum over
the coefficients $c_n^{(s-1)}$ of the prior, but without any other infinite series.

\subsection{Simplifying and solving the series}
Using some results from \cite{2012boyadzhiev}, we find the following series solutions
\begin{align}
  \sum_{n=1}^\infty \frac{1}{n(2n-1)}
  \begin{pmatrix}
    2n\\
    n
  \end{pmatrix}
  x^n &= \frac{8x}{1 + \sqrt{1-4x}} + 2\ln(1+\sqrt{1-4x}) - 2\ln(2) \, ,
        \label{eq:bin-series-A}
\end{align}
\begin{align}
  \sum_{n=m}^\infty \frac{1}{n(2n-1)}
  \begin{pmatrix}
    2n\\
    n-m
  \end{pmatrix}
  x^n &= \left( \frac{1+2m\sqrt{1-4x}}{m(4m^2-1)} \right)\left( \frac{4x}{(1+\sqrt{1-4x})^2} \right)^m \, ,
        \label{eq:bin-series-B}
\end{align}
\begin{align}
  &\sum_{v=m}^\infty \left(\frac{1}{2(v-1)(2v-1)}\right)
  \begin{pmatrix}
    2v - 1 \\
    v - m
  \end{pmatrix}
  x^{2v - 1} \nonumber \\
  &= \frac{x}{2} \left( \frac{1+(2m-1)\sqrt{1-4x^2}}{(m-1)m(2m-1)(1+\sqrt{1-4x^2})} \right)
    \left( \frac{4x^2}{(1+\sqrt{1-4x^2})^2} \right)^{m-1} \, ,
    \label{eq:bin-series-C}
\end{align}
\begin{align}
  &\sum_{v=2}^\infty \frac{1}{2(v-1)(2v-1)}
    \begin{pmatrix}
      2v-1 \\
      v-1
    \end{pmatrix} x^{2v-1} \nonumber \\
  &= \frac{x}{2}
    \left( 1 + 2\ln(2) - 2\ln(1 + \sqrt{1-4x^2}) - \frac{2}{1 + \sqrt{1-4x^2}} \right) \, ,
    \label{eq:bin-series-D}
\end{align}
which will be used in the derivation below.

We rewrite the sum over $q$ in \eqref{eq:int-sol} separately for the cases where $u$ is
even or odd. If $u$ is even, we have
\begin{align*}
  \sum_{q=0}^u
  \begin{pmatrix}
    u \\
    q
  \end{pmatrix}
  \Exp{i\alpha(u-2q)} c_{k(u-2q)}^{(s-1)}
  &= \sum_{m=-u/2}^{u/2}
    \begin{pmatrix}
      u \\
      \frac{u}{2} - m
    \end{pmatrix}
  \Exp{i\alpha 2m} c_{k 2m}^{(s-1)} \, .
\end{align*}
For every term in the sum with a positive value of $m$ there is another term with the same
value of $m$, but negative; and there is one term with $m=0$. So we can rewrite the sum as
\begin{align*}
  \begin{pmatrix}
    u \\
    \frac{u}{2}
  \end{pmatrix} c_0^{(s-1)} +
  \sum_{m=1}^{u/2} \Bigg(
  \begin{pmatrix}
    u \\
    \frac{u}{2} - m
  \end{pmatrix}
  \Exp{i\alpha 2m} c_{k 2m}^{(s-1)} +
  \begin{pmatrix}
    u \\
    \frac{u}{2} + m
  \end{pmatrix}
  \Exp{-i\alpha 2m} c_{-k 2m}^{(s-1)}\Bigg) \, .
\end{align*}
We can use the property of binomial coefficients:
\begin{align*}
  \begin{pmatrix}
    n \\
    k
  \end{pmatrix} &=
                  \begin{pmatrix}
                    n \\
                    n-k
                  \end{pmatrix}
  \text{, so we have }
  \begin{pmatrix}
    u \\
    \frac{u}{2} - m
  \end{pmatrix} =
  \begin{pmatrix}
    u \\
    \frac{u}{2} + m
  \end{pmatrix} \, .
\end{align*}
Also noting that $c_0^{(s-1)} = 1$ \footnote{This is required by normalisation of
  $p_{s-1}(\phi)$.} we can rewrite the sum as
\begin{align*}
  \begin{pmatrix}
    u \\
    \frac{u}{2}
  \end{pmatrix}
  + 2\sum_{m=1}^{u/2}
  \begin{pmatrix}
    u \\
    \frac{u}{2} - m
  \end{pmatrix}
  \mathfrak{Re} \Big\{ \Exp{i\alpha 2m} c_{k 2m}^{(s-1)} \Big\} \, .
\end{align*}
Similarly, if $u$ is odd, we can write
\begin{align*}
  \sum_{q=0}^u
  \begin{pmatrix}
    u \\
    q
  \end{pmatrix}
  \Exp{i\alpha(u-2q)} c_{k(u-2q)}^{(s-1)}
  &= 2 \sum_{m=1}^{(u+1)/2}
    \begin{pmatrix}
      u \\
      \frac{u + 1}{2} - m
    \end{pmatrix}
  \mathfrak{Re} \Big\{ \Exp{i\alpha(2m-1)} c_{k(2m-1)}^{(s-1)} \Big\} \, .
\end{align*}
Now let
\begin{align*}
  f(u) \equiv \sum_{q=0}^u
  \begin{pmatrix}
    u \\
    q
  \end{pmatrix}
  \Exp{i\alpha(u-2q)} c_{k(u-2q)}^{(s-1)} \, ,
\end{align*}
to simplify the notation. So splitting the sum over $u$ into even and odd parts, we have
\begin{align*}
  \sum_{u=0}^{2n}
  \begin{pmatrix}
    2n \\
    u
  \end{pmatrix}
  (1-\lambda)^{2n-u} \left(\frac{\lambda \zeta}{2}\right)^u f(u)
  &= \sum_{u=0}^{n}
    \begin{pmatrix}
      2n \\
      2u
    \end{pmatrix}
  (1-\lambda)^{2(n-u)} \left(\frac{\lambda \zeta}{2}\right)^{2u} f(2u) \\
  &+ \sum_{v=1}^{n}
    \begin{pmatrix}
      2n \\
      2v - 1
    \end{pmatrix}
  (1-\lambda)^{2(n-v) + 1} \left(\frac{\lambda \zeta}{2}\right)^{2v-1} f(2v-1) \, .
\end{align*}
Now we consider the even part of the sum:
\begin{align*}
  &\sum_{u=0}^{n}
    \begin{pmatrix}
      2n \\
      2u
    \end{pmatrix}
  (1-\lambda)^{2(n-u)} \left(\frac{\lambda \zeta}{2}\right)^{2u} f(2u) \\
  &= (1-\lambda)^{2n} + \sum_{u=1}^{n}
    \begin{pmatrix}
      2n \\
      2u
    \end{pmatrix}
  (1-\lambda)^{2(n-u)} \left(\frac{\lambda \zeta}{2}\right)^{2u}
  \begin{pmatrix}
    2u \\
    u
  \end{pmatrix} \\
  &+ 2 \sum_{u=1}^{n}
    \begin{pmatrix}
      2n \\
      2u
    \end{pmatrix}
  (1-\lambda)^{2(n-u)} \left(\frac{\lambda \zeta}{2}\right)^{2u} \sum_{m=1}^{u}
  \begin{pmatrix}
    2u \\
    u - m
  \end{pmatrix}
  \mathfrak{Re} \Big\{ \Exp{i\alpha 2m} c_{k 2m}^{(s-1)} \Big\} \, .
\end{align*}
For the odd part we have
\begin{align*}
  &\sum_{v=1}^{n}
    \begin{pmatrix}
      2n \\
      2v - 1
    \end{pmatrix}
  (1-\lambda)^{2(n-v) + 1} \left(\frac{\lambda \zeta}{2}\right)^{2v-1} f(2v-1) \\
  &= 2\sum_{v=1}^{n}
    \begin{pmatrix}
      2n \\
      2v - 1
    \end{pmatrix}
  (1-\lambda)^{2(n-v) + 1} \left(\frac{\lambda \zeta}{2}\right)^{2v-1}
  \sum_{m=1}^v
  \begin{pmatrix}
    2v-1 \\
    v - m
  \end{pmatrix}
  \mathfrak{Re} \Big\{ \Exp{i\alpha(2m-1)} c_{k(2m-1)}^{(s-1)} \Big\} \, .
\end{align*}
So if we combine everything we have
\begin{align}
  &\sum_{\xi} \int_0^{2\pi} \frac{d\phi}{2\pi} P_{\xi}(\alpha , k\phi)p_{s-1}(\phi)
    \ln \big(P_{\xi}(\alpha , k\phi)\big) \nonumber \\
  &= -\ln(2) + \sum_{n=1}^\infty \left( \frac{1}{2n(2n-1)} \right) \Bigg[
    (1-\lambda)^{2n} + \sum_{u=1}^{n}
    \begin{pmatrix}
      2n \\
      2u
    \end{pmatrix}
  (1-\lambda)^{2(n-u)} \left(\frac{\lambda \zeta}{2}\right)^{2u}
  \begin{pmatrix}
    2u \\
    u
  \end{pmatrix} \nonumber \\
  &+ 2 \sum_{u=1}^{n}
    \begin{pmatrix}
      2n \\
      2u
    \end{pmatrix}
  (1-\lambda)^{2(n-u)} \left(\frac{\lambda \zeta}{2}\right)^{2u} \sum_{m=1}^{u}
  \begin{pmatrix}
    2u \\
    u - m
  \end{pmatrix}
  \mathfrak{Re} \Big\{ \Exp{i\alpha 2m} c_{k 2m}^{(s-1)} \Big\} \nonumber \\
  &+ 2\sum_{v=1}^{n}
    \begin{pmatrix}
      2n \\
      2v - 1
    \end{pmatrix}
  (1-\lambda)^{2(n-v) + 1} \left(\frac{\lambda \zeta}{2}\right)^{2v-1}
  \sum_{m=1}^v
  \begin{pmatrix}
    2v-1 \\
    v - m
  \end{pmatrix}
  \mathfrak{Re} \Big\{ \Exp{i\alpha(2m-1)} c_{k(2m-1)}^{(s-1)} \Big\}
  \Bigg] \, . \label{eq:4-series}
\end{align}
There are now four terms in the infinite sum over $n$. The first is
\begin{align}
  \sum_{n=1}^\infty \left( \frac{1}{2n(2n-1)} \right) (1-\lambda)^{2n}
  &= \frac{1}{2} \ln\left( 1 - (1-\lambda)^2 \right) + (1-\lambda)\tanh^{-1}(1-\lambda) \, .
    \label{eq:1st-term-sol}
\end{align}
The second is
\begin{align*}
  &\sum_{n=1}^\infty \left( \frac{1}{2n(2n-1)} \right) \sum_{u=1}^{n}
  \begin{pmatrix}
    2n \\
    2u
  \end{pmatrix}
  (1-\lambda)^{2(n-u)} \left(\frac{\lambda \zeta}{2}\right)^{2u}
  \begin{pmatrix}
    2u \\
    u
  \end{pmatrix} \\
  &= \sum_{m=1}^\infty \left(\frac{\lambda \zeta}{2}\right)^{2m}
    \begin{pmatrix}
      2m \\
      m
    \end{pmatrix}
  (1-\lambda)^{-2m}
  \sum_{n=m}^\infty \left( \frac{1}{2n(2n-1)} \right)
  \begin{pmatrix}
    2n \\
    2m
  \end{pmatrix}
  (1-\lambda)^{2n} \, .
\end{align*}
Using the solution to the infinite series:
\begin{align}
  \sum_{n=m}^\infty \frac{x^{2n}}{2n(2n-1)}
  \begin{pmatrix}
    2n \\
    2m
  \end{pmatrix}
  &= \frac{x^{2m}}{4m(2m-1)} \Bigg(
    \frac{(1+x)}{(1+x)^{2m}} + \frac{(1-x)}{(1-x)^{2m}}
    \Bigg) \, , \label{eq:series-2n-2m}
\end{align}
we can rewrite the second term as
\begin{align*}
    &\frac{(2-\lambda)}{4} \sum_{m=1}^\infty \left(\frac{1}{m(2m-1)}\right)
      \begin{pmatrix}
        2m \\
        m
      \end{pmatrix}
  \left(\frac{\lambda \zeta}{2(2-\lambda)}\right)^{2m} \\
    &+ \frac{\lambda}{4} \sum_{m=1}^\infty \left(\frac{1}{m(2m-1)}\right)
      \begin{pmatrix}
        2m \\
        m
      \end{pmatrix}
  \left(\frac{\zeta}{2}\right)^{2m} \, .
\end{align*}
Using the result \eqref{eq:bin-series-A} we find this is equal to
\begin{align}
  \left(1 - \frac{\lambda}{2}\right)F(\delta) + \frac{\lambda}{2}F(\zeta) - \ln(2) \, ,
  \label{eq:2nd-term-sol}
\end{align}
where we defined
\begin{align}
  \delta \equiv \frac{\lambda\zeta}{2 - \lambda} , \qquad
  F(x) \equiv \frac{x^2}{g_1(x)} + \ln\left(g_1(x)\right) , \qquad
  g_1(x) \equiv 1 + g_0(x) , \qquad
  g_0(x) \equiv \sqrt{1 - x^2} \, .
  \label{eq:define-delta-F-g12}
\end{align}
Next we consider the third term in the infinite sum in \eqref{eq:4-series}:
\begin{align*}
  &\sum_{n=1}^\infty \left( \frac{1}{n(2n-1)} \right) \sum_{u=1}^{n}
    \begin{pmatrix}
      2n \\
      2u
    \end{pmatrix}
  (1-\lambda)^{2(n-u)} \left(\frac{\lambda \zeta}{2}\right)^{2u} \sum_{m=1}^{u}
  \begin{pmatrix}
    2u \\
    u - m
  \end{pmatrix}
  \mathfrak{Re} \Big\{ \Exp{i\alpha 2m} c_{k 2m}^{(s-1)} \Big\} \\
  &= \sum_{m=1}^\infty
    \mathfrak{Re} \Big\{ \Exp{i\alpha 2m} c_{k 2m}^{(s-1)} \Big\}
    \sum_{u=m}^\infty \left(\frac{\lambda \zeta}{2}\right)^{2u}
    \begin{pmatrix}
      2u \\
      u - m
    \end{pmatrix} (1-\lambda)^{-2u}
  \sum_{n=u}^\infty \left( \frac{1}{n(2n-1)} \right)
  \begin{pmatrix}
    2n \\
    2u
  \end{pmatrix} (1-\lambda)^{2n} \, .
\end{align*}
The last series over $n$ is given by \eqref{eq:series-2n-2m} (within a factor of two), so we have
\begin{align*}
  &\sum_{u=m}^\infty \left(\frac{\lambda \zeta}{2}\right)^{2u}
    \begin{pmatrix}
      2u \\
      u - m
    \end{pmatrix} (1-\lambda)^{-2u}
  \sum_{n=u}^\infty \left( \frac{1}{n(2n-1)} \right)
  \begin{pmatrix}
    2n \\
    2u
  \end{pmatrix} (1-\lambda)^{2n} \\
  &= \frac{(2-\lambda)}{2} \sum_{u=m}^\infty
    \left(\frac{1}{u(2u-1)}\right)
    \begin{pmatrix}
      2u \\
      u - m
    \end{pmatrix}
  \left(\frac{\lambda \zeta}{2(2-\lambda)}\right)^{2u} \\
  &+ \frac{\lambda}{2} \sum_{u=m}^\infty \left(\frac{1}{u(2u-1)}\right)
    \begin{pmatrix}
      2u \\
      u - m
    \end{pmatrix}
  \left(\frac{\zeta}{2}\right)^{2u} \, .
\end{align*}
Using \eqref{eq:bin-series-B} we find this is equal to
\begin{align}
  \left(1 - \frac{\lambda}{2}\right)G(\delta, m) + \frac{\lambda}{2}G(\zeta, m) \, ,
  \label{eq:3rd-term-sol}
\end{align}
where we defined
\begin{align}
  G(x, m) \equiv \left(\frac{1 + 2m g_0(x)}{m(4m^2 - 1)}\right)\left(\frac{x}{g_1(x)}\right)^{2m}
  \, . \label{eq:define-G}
\end{align}
Last we consider the fourth term in the infinite sum in \eqref{eq:4-series}:
\begin{align*}
  &\sum_{n=1}^\infty \left( \frac{1}{n(2n-1)} \right) \sum_{v=1}^{n}
  \begin{pmatrix}
    2n \\
    2v - 1
  \end{pmatrix}
  (1-\lambda)^{2(n-v) + 1} \left(\frac{\lambda \zeta}{2}\right)^{2v-1}
  \sum_{m=1}^v
  \begin{pmatrix}
    2v-1 \\
    v - m
  \end{pmatrix}
  \mathfrak{Re} \Big\{ \Exp{i\alpha(2m-1)} c_{k(2m-1)}^{(s-1)} \Big\} \\
  &= \sum_{m=1}^\infty \mathfrak{Re} \Big\{ \Exp{i\alpha(2m-1)} c_{k(2m-1)}^{(s-1)} \Big\}
    \sum_{v=m}^\infty
    \begin{pmatrix}
      2v-1 \\
      v - m
    \end{pmatrix}
  \left(\frac{\lambda \zeta}{2}\right)^{2v-1} \\
  &\qquad (1-\lambda)^{-(2v - 1)}
    \sum_{n=v}^\infty \left( \frac{1}{n(2n-1)} \right)
    \begin{pmatrix}
      2n \\
      2v - 1
    \end{pmatrix}
  (1-\lambda)^{2n} \, .
\end{align*}
Here the last series over $n$ is given by (for $m > 1$; we will need to consider the
$m = 1$ case separately)
\begin{align*}
  \sum_{n=m}^\infty \left(\frac{x^{2n}}{n(2n-1)}\right)
  \begin{pmatrix}
    2n \\
    2m - 1
  \end{pmatrix}
  &= \left(\frac{x^{2m-1}}{2(m-1)(2m-1)}\right)
    \left(\frac{(1-x)}{(1-x)^{2m-1}}-\frac{(1+x)}{(1+x)^{2m-1}}\right) \, ,
\end{align*}
and for $m = 1$
\begin{align*}
  &\sum_{n=m}^\infty \left(\frac{x^{2n}}{n(2n-1)}\right)
  \begin{pmatrix}
    2n \\
    2m - 1
  \end{pmatrix}
  = \sum_{n=1}^\infty \frac{x^{2n}}{n(2n-1)} 2n
  = 2\sum_{n=1}^\infty \frac{x^{2n}}{(2n-1)}
  = 2x \tanh^{-1} (x) \, ,
\end{align*}
so we have (for $m > 1$)
\begin{align*}
  &\sum_{v=m}^\infty
    \begin{pmatrix}
      2v-1 \\
      v - m
    \end{pmatrix}
  \left(\frac{\lambda \zeta}{2}\right)^{2v-1}
  (1-\lambda)^{-(2v - 1)}
  \sum_{n=v}^\infty \left( \frac{1}{n(2n-1)} \right)
  \begin{pmatrix}
    2n \\
    2v - 1
  \end{pmatrix}
  (1-\lambda)^{2n} \\
  &= \lambda \sum_{v=m}^\infty
    \left(\frac{1}{2(v-1)(2v-1)}\right)
    \begin{pmatrix}
      2v-1 \\
      v - m
    \end{pmatrix}
  \left(\frac{\zeta}{2}\right)^{2v-1} \\
  &- (2-\lambda) \sum_{v=m}^\infty
    \left(\frac{1}{2(v-1)(2v-1)}\right)
    \begin{pmatrix}
      2v-1 \\
      v - m
    \end{pmatrix}
  \left(\frac{\lambda \zeta}{2(2-\lambda)}\right)^{2v-1} \, .
\end{align*}
Using \eqref{eq:bin-series-C} we find this is equal to
\begin{align}
  \frac{\lambda\zeta}{4}
  \left(J(\zeta, m) - J(\delta, m)\right) \, ,
  \label{eq:4th-term-m>1-sol}
\end{align}
where we defined
\begin{align}
  J(x, m) \equiv \left(\frac{1 + (2m-1)g_0(x)}{(m-1)m(2m-1)g_1(x)}\right)
  \left(\frac{x}{g_1(x)}\right)^{2(m-1)} \, .
  \label{eq:define-J}
\end{align}
Finally, for the $m = 1$ case in the fourth term in \eqref{eq:4-series}, we have
\begin{align*}
  &\sum_{v=m}^\infty
    \begin{pmatrix}
      2v-1 \\
      v - m
    \end{pmatrix}
  \left(\frac{\lambda \zeta}{2}\right)^{2v-1}
  (1-\lambda)^{-(2v - 1)}
  \sum_{n=v}^\infty \left( \frac{1}{n(2n-1)} \right)
  \begin{pmatrix}
    2n \\
    2v - 1
  \end{pmatrix}
  (1-\lambda)^{2n} \\
  &= \lambda \zeta \tanh^{-1}(1-\lambda) \\
  &+ \sum_{v=2}^\infty
    \left(\frac{1}{2(v-1)(2v-1)}\right)
    \begin{pmatrix}
      2v-1 \\
      v - 1
    \end{pmatrix}
  \left(\lambda \left(\frac{\zeta}{2}\right)^{2v-1} - (2-\lambda) \left(\frac{\lambda \zeta}{2(2-\lambda)}\right)^{2v-1}\right) \, .
\end{align*}
Using \eqref{eq:bin-series-D} we can rewrite this as
\begin{align}
  \lambda\zeta\left(\tanh^{-1}(1-\lambda) + \frac{L(\delta) - L(\zeta)}{2}\right) \, ,
  \label{eq:4th-term-m=1-sol}
\end{align}
where we defined
\begin{align}
  L(x) \equiv \frac{1}{g_1(x)} + \ln\left(g_1(x)\right)
  \label{eq:define-L}
\end{align}

\subsection{Summary}
Substituting the results and definitions \eqref{eq:1st-term-sol}, \eqref{eq:2nd-term-sol},
\eqref{eq:define-delta-F-g12}, \eqref{eq:3rd-term-sol}, \eqref{eq:define-G},
\eqref{eq:4th-term-m>1-sol}, \eqref{eq:define-J}, \eqref{eq:4th-term-m=1-sol}, and
\eqref{eq:define-L} into \eqref{eq:4-series}, we have
\begin{align*}
  &\sum_{\xi} \int_0^{2\pi} \frac{d\phi}{2\pi} P_{\xi}(\alpha , k\phi)p_{s-1}(\phi)
    \ln \big(P_{\xi}(\alpha , k\phi)\big) \\
  &= -\ln(2) + \frac{1}{2} \ln\left( 1 - (1-\lambda)^2 \right) + (1-\lambda)\tanh^{-1}(1-\lambda)
  + \left(1 - \frac{\lambda}{2}\right) F(\delta) + \frac{\lambda}{2} F(\zeta) - \ln(2) \\
  &+ \sum_{m=1}^\infty A_m \mathfrak{Re} \left\{ \Exp{i\alpha 2m} c_{k2m}^{(s-1)} \right\}
    + B_m \mathfrak{Re} \left\{ \Exp{i\alpha (2m-1)} c_{k(2m-1)}^{(s-1)} \right\} \, ,
\end{align*}
where
\begin{align*}
  A_m &= \left(1 - \frac{\lambda}{2}\right) G(\delta, m)
        + \frac{\lambda}{2} G(\zeta, m) \\
  B_m &=
        \begin{cases}
          \frac{\lambda\zeta}{2}
          \left(\ln\left(1 - \frac{\lambda}{2}\right) + L(\delta)
          - \ln\left(\frac{\lambda}{2}\right) - L(\zeta)\right)
        \, , &\text{if } m = 1 \\
        \frac{\lambda\zeta}{4} \left(J(\zeta, m) - J(\delta, m)\right)
        \, , &\text{otherwise}
        \end{cases}
\end{align*}
and the expected entropy gain \eqref{eq:entropy-gain} is
\begin{align}
  \Delta_s H(\alpha,\, k) &= -2\ln(2) + \frac{1}{2} \ln\left( 1 - (1-\lambda)^2 \right)
                            + \frac{(1-\lambda)}{2}\ln\left(\frac{2-\lambda}{\lambda}\right)
  + \left(1 - \frac{\lambda}{2}\right) F(\delta) + \frac{\lambda}{2} F(\zeta) \nonumber \\
                          &+ \sum_{m=1}^\infty \Big( A_m \mathfrak{Re} \left\{ \Exp{i\alpha 2m} c_{k2m}^{(s-1)} \right\}
                            + B_m \mathfrak{Re} \left\{ \Exp{i\alpha (2m-1)} c_{k(2m-1)}^{(s-1)} \right\} \Big)
                            - \sum_{\xi} \Pi_{\xi}(\alpha,\, k) \ln \left(\Pi_{\xi}(\alpha,\, k)\right) \, . \label{eq:entropy-gain-SM}
\end{align}
\subsection{Limiting cases}
If $\lambda \rightarrow 1$, then $\delta \rightarrow \zeta$, and we
have
\begin{align*}
  \Delta_s H(\alpha,\, k) &= -2\ln(2) + \frac{1}{2} \ln(1) + (0)\ln(1) + F(\zeta) \\
                          &+ \sum_{m=1}^\infty \Big( A_m \mathfrak{Re} \left\{ \Exp{i\alpha 2m} c_{k2m}^{(s-1)} \right\}
                            + B_m \mathfrak{Re} \left\{ \Exp{i\alpha (2m-1)} c_{k(2m-1)}^{(s-1)} \right\} \Big) \\
                          &- \sum_{\xi} \Pi_{\xi}(\alpha,\, k) \ln \left(\Pi_{\xi}(\alpha,\, k)\right) \, ,
\end{align*}
and $A_m = G(\zeta, m), \, B_m = 0$, so
\begin{align*}
  \Delta_s H(\alpha,\, k) &= -2\ln(2) + F(\zeta)
                            + \sum_{m=1}^\infty G(\zeta, m)
                            \mathfrak{Re} \left\{ \Exp{i\alpha 2m} c_{k2m}^{(s-1)} \right\}
                            - \sum_{\xi} \Pi_{\xi}(\alpha,\, k) \ln \left(\Pi_{\xi}(\alpha,\, k)\right) \, .
\end{align*}
If we also have $\zeta = 1$, then $g_0(\zeta) = 0$, and we have
\begin{align*}
  \Delta_s H(\alpha,\, k) &= -2\ln(2) + 1 + \sum_{m=1}^\infty \left(\frac{1}{m(4m^2 -1)}\right)
                            \mathfrak{Re} \left\{ \Exp{i\alpha 2m} c_{k2m}^{(s-1)} \right\}
                            - \sum_{\xi} \Pi_{\xi}(\alpha,\, k) \ln \left(\Pi_{\xi}(\alpha,\, k)\right) \, .
\end{align*}
If $\zeta = 0$, then $g_0(\zeta) = 1$ and we have
\begin{align*}
  \Delta_s H(\alpha,\, k) &= -2\ln(2) + \ln(2)
                            - \sum_{\xi} \Pi_{\xi}(\alpha,\, k) \ln \left(\Pi_{\xi}(\alpha,\, k)\right) \, .
\end{align*}
Since in this case $\Pi_{\xi}(\alpha,\, k) = \frac{1}{2}$, this is
\begin{align*}
  \Delta_s H(\alpha,\, k) &= -\ln(2) - \ln\left(\frac{1}{2}\right) = 0 \, ,
\end{align*}
as expected, since no information about $\phi$ is given by the measurement.

Finally, we consider the case $\lambda \rightarrow 0$. Then $\delta \rightarrow 0$, and we
have
\begin{align*}
  \Delta_s H(\alpha,\, k) &= -2\ln(2) + \frac{1}{2} \ln(0)
                            + \frac{1}{2}\ln\left(\frac{2}{0}\right) + \ln(2) \\
                          &+ \sum_{m=1}^\infty \Big( A_m \mathfrak{Re} \left\{ \Exp{i\alpha 2m} c_{k2m}^{(s-1)} \right\}
                            + B_m \mathfrak{Re} \left\{ \Exp{i\alpha (2m-1)} c_{k(2m-1)}^{(s-1)} \right\} \Big) \\
                          &- \sum_{\xi} \Pi_{\xi}(\alpha,\, k) \ln \left(\Pi_{\xi}(\alpha,\, k)\right) \, ,
\end{align*}
and
\begin{align*}
  A_m &= 0 \\
  B_m &=
        \begin{cases}
          0 \ln\left(\frac{2}{0}\right)
          \, , &\text{if } m = 1 \\
          0 \, , &\text{otherwise}
        \end{cases} \, ,
\end{align*}
and since $\Pi_{\xi}(\alpha,\, k) \in \{1,0\}$,
\begin{align*}
  \sum_{\xi} \Pi_{\xi}(\alpha,\, k) \ln \left(\Pi_{\xi}(\alpha,\, k)\right) = 0 \, .
\end{align*}
But we see that the value of
\begin{align*}
  \frac{1}{2} \ln\left( 1 - (1-\lambda)^2 \right)
  + \frac{(1-\lambda)}{2}\ln\left(\frac{2-\lambda}{\lambda}\right)
  &\rightarrow \frac{1}{2} \ln(0) + \frac{1}{2} \ln \left(\frac{2}{0}\right)
\end{align*}
is undefined. Recalling that this was the result of the series
\begin{align*}
  \sum_{n=1}^\infty \left( \frac{1}{2n(2n-1)} \right) (1-\lambda)^{2n} \, ,
\end{align*}
we can see that if we set $\lambda = 0$, we have
\begin{align*}
  \sum_{n=1}^\infty \left( \frac{1}{2n(2n-1)} \right) (1-\lambda)^{2n} &\rightarrow \sum_{n=1}^\infty \left( \frac{1}{2n(2n-1)} \right) \\
                                                                       &= \ln(2) \, .
\end{align*}
Similarly, the value of $B_1$ is undefined. We see that all terms in $B_1$ are zero,
except one, which becomes undefined at $\lambda = 0$; this is the term that results from
the series:
\begin{align*}
  \lambda \zeta (1-\lambda)^{-1}
  \sum_{n=1}^\infty \left( \frac{1}{(2n-1)} \right) (1-\lambda)^{2n}
  &\rightarrow (0) \zeta \sum_{n=1}^\infty \frac{1}{2n-1} \, .
\end{align*}
But since the series term diverges, we need to take the limit:
\begin{align*}
  \lim_{\lambda \rightarrow 0} \lambda \ln\left(\frac{2-\lambda}{\lambda}\right) &= 0 \, .
\end{align*}
So $B_1 \rightarrow 0$. Putting these results back together, we have
\begin{align*}
  \Delta_s H(\alpha,\, k) &= -2\ln(2) + \ln(2) + \ln(2) = 0 \, .
\end{align*}
Similarly to the $\zeta = 0$ case, when $\lambda = 0$ the measurement gives no information
about $\phi$ so the expected entropy gain is zero.

\section{\label{sec:sup-sh-gain} Expected sharpness gain}

The sharpness at step $s$ is
\begin{align*}
  S\left[p_s(\phi)\right]
  &= \left| \expect{\Exp{i\phi}}_s \right|
    = \left| \int_0^{2\pi} \frac{d\phi}{2\pi} p_s(\phi) \Exp{i\phi} \right|
    = \left| c_{-1}^{(s)} \right|
\end{align*}
We define the expected sharpness gain for step $s$:
\begin{align*}
  \Delta_s S(\alpha,\, k)
  &= \sum_{\xi} \Pi_{\xi}(\alpha,\, k)
    \Big( S\left[ p_s(\phi|\xi ; \alpha, k) \right]
    - S\left[ p_{s-1}(\phi) \right] \Big) \\
  &= \sum_{\xi} \Pi_{\xi}(\alpha,\, k)
    S\left[ p_s(\phi|\xi ; \alpha, k) \right]
    - \left| c_{-1}^{(s-1)} \right| \\
  &= \sum_{\xi}
    \left| \int_0^{2\pi} \frac{d\phi}{2\pi} P_{\xi}(\alpha, k\phi)p_{s-1}(\phi) \Exp{i\phi} \right|
    - \left| c_{-1}^{(s-1)} \right| \, .
\end{align*}
We have
\begin{align}
  &\int_0^{2\pi} \frac{d\phi}{2\pi} P_{\xi}(\alpha, k\phi)p_{s-1}(\phi) \Exp{i\phi} \nonumber \\
  &= \int_0^{2\pi} \frac{d\phi}{2\pi}
    \sum_{n = -\infty}^\infty \Bigg[
    \frac{1}{2}\big(1 + \xi(1 - \lambda)\big) c_n^{(s-1)}
    + \xi\lambda\frac{\zeta}{4} \left( \Exp{i\alpha}c_{n+k}^{(s-1)} + \Exp{-i\alpha}c_{n-k}^{(s-1)} \right)
    \Bigg] \Exp{i n\phi} \Exp{i\phi} \nonumber \\
  &= \sum_{n = -\infty}^\infty \Bigg[
    \frac{1}{2}\big(1 + \xi(1 - \lambda)\big) c_n^{(s-1)}
    + \xi\lambda\frac{\zeta}{4} \left( \Exp{i\alpha}c_{n+k}^{(s-1)} + \Exp{-i\alpha}c_{n-k}^{(s-1)} \right)
    \Bigg] \int_0^{2\pi} \frac{d\phi}{2\pi} \Exp{i (n+1) \phi} \nonumber \\
  &= \frac{1}{2}\big(1 + \xi(1 - \lambda)\big) c_{-1}^{(s-1)}
    + \xi\lambda\frac{\zeta}{4} \left( \Exp{i\alpha}c_{-1+k}^{(s-1)} + \Exp{-i\alpha}c_{-1-k}^{(s-1)} \right) \, ,
    \label{eq:post-sharp}
\end{align}
Using \eqref{eq:post-sharp}, we have
\begin{align}
  \Delta_s S(\alpha,\, k)
  &= \sum_{\xi} \bigg|
    \frac{1}{2}\big(1 + \xi(1 - \lambda)\big) c_{-1}^{(s-1)}
    + \xi\lambda\frac{\zeta}{4} \left( \Exp{i\alpha}c_{-1+k}^{(s-1)} + \Exp{-i\alpha}c_{-1-k}^{(s-1)} \right)
    \bigg| - \left| c_{-1}^{(s-1)} \right| \, . \label{eq:sharpness-gain-SM}
\end{align}

\subsection{Limiting cases}
If $\lambda = 1$
\begin{align*}
  \Delta_s S(\alpha,\, k)
  &= \sum_{\xi} \bigg|
    \frac{1}{2} c_{-1}^{(s-1)}
    + \xi\frac{\zeta}{4} \left( \Exp{i\alpha}c_{-1+k}^{(s-1)} + \Exp{-i\alpha}c_{-1-k}^{(s-1)} \right)
    \bigg| - \left| c_{-1}^{(s-1)} \right| \, ,
\end{align*}
and if $\zeta = 0$
\begin{align*}
  \Delta_s S(\alpha,\, k)
  &= \sum_{\xi} \bigg| \frac{1}{2} c_{-1}^{(s-1)} \bigg|
    - \left| c_{-1}^{(s-1)} \right|
    = \frac{1}{2} 2\bigg| c_{-1}^{(s-1)} \bigg|
    - \left| c_{-1}^{(s-1)} \right| = 0 \, ,
\end{align*}
as expected. And if $\lambda = 0$
\begin{align*}
  \Delta_s S(\alpha,\, k)
  &= \sum_{\xi} \bigg| \frac{1}{2}\big(1 + \xi\big) c_{-1}^{(s-1)} \bigg|
    - \left| c_{-1}^{(s-1)} \right|
    = \bigg| \frac{1}{2}2 c_{-1}^{(s-1)} \bigg| - \left| c_{-1}^{(s-1)} \right| = 0 \, ,
\end{align*}
as expected.

\end{document}